\documentclass[preprint,12pt,sort&compress]{elsarticle}

\usepackage[reviewcopy]{adndt}

\journal{Atomic Data Nuclear Data Tables}

\usepackage{array}
\usepackage{amssymb}
\usepackage{amsmath}
\usepackage{caption}
\usepackage{comment}
\usepackage{float}
\usepackage{graphicx}
\usepackage[colorlinks]{hyperref}
\usepackage{cleveref}
\usepackage{isotope}
\usepackage{lineno}
\usepackage{longtable}
\usepackage{multirow}
\usepackage{makecell}
\usepackage[version=4]{mhchem}
\usepackage{natbib}
\usepackage{subcaption}
\usepackage{tablefootnote}
\usepackage{tabulary}
\usepackage{threeparttable}

\begin{document}

\begin{frontmatter}

\title{The Confined \(\beta \)-Soft rotor model in rare-earth nuclei}

\author[first]{Jim A. Papadopoulos\fnmark[1]}
\author[first]{T.J. Mertzimekis}
\author[first]{P. Koseoglou}
\author[first]{P. Vasileiou\fnmark[2]}
\author[second]{Dennis Bonatsos}

\affiliation[first]{organization={National and Kapodistrian University of Athens},
            addressline={Zografou Campus}, 
            city={Athens},
            postcode={GR-15784}, 
            country={Greece}}

\affiliation[second]{
    organization={Institute of Nuclear and Particle Physics, National Centre for Scientific Research ``Demokritos''},
    addressline={Aghia Paraskevi},
    city={Athens},
    postcode={GR-15310},
    country={Greece}
}

\fntext[1]{Present address: Department of Physics, University of Guelph, Guelph, Ontario N1G 2W1, Canada}

\fntext[2]{Present address: Horia Hulubei National Institute of Physics and Nuclear Engineering - IFIN-HH, R-077125 Bucharest, Romania}
\begin{abstract}

Contemporary theoretical descriptions of nuclear structure rely mainly on
microscopic, single-particle frameworks often in competition with collective degrees of freedom, especially when deformation plays a dominant role. Such phenomena are
prominent in the rare-earth region, where rotational band structures and enhanced
electric quadrupole transitions are systematically examined. The Confined \(\beta \)-Soft (CBS) rotor
model, introduced by N. Pietralla and O.M. Gorbachenko, bridges the gap between the X(5)
critical point and the rigid-rotor limit in the region where the  
$R_{4/2}=E(4^+)/E(2^+)$ ratio lies between 2.904 and 3.333.
In the present work, the CBS framework is employed to calculate ground-state band 
energies, associated B(E2) transition rates, and \(\beta \)-band excitations of
even-even nuclei in the rare-earth region. The theoretical results are systematically 
compared with available experimental data, and predictions are provided for nuclear observables that have not yet been measured, offering guidance for future
experimental investigations.

\end{abstract}



\begin{keyword}
CBS model\sep
rare-earth \sep
rigid-rotor \sep
deformation



\end{keyword}

\end{frontmatter}

\clearpage

\tableofcontents
\listofDtables
\listofDfigures

\clearpage

\section{Introduction}
\label{sec:intro}

Structural characteristics in atomic nuclei are almost always not possible to be explained
by the simple picture of a spherical nucleus. Indeed, in many regions of the nuclear
chart, particularly among medium- and heavy-mass nuclei, an increasing number of valence
nucleons leads to spontaneous breaking of spherical symmetry, resulting in deformed 
shapes~\cite{Mottelson_2012}. This nuclear deformation has profound consequences for
the structure and excitation modes of nuclei and also introduces considerable complexity
in the theoretical description and prediction of the evolution of the nuclear structure.
Deformed nuclei are typically associated with collective phenomena, where nucleons move
coherently, giving rise to rotational bands and vibrational excitations that differ
markedly from those found in spherical systems. Although collectivity has been
extensively observed and studied for decades, its precise theoretical treatment remains
an open challenge~\cite{Litvinova2025NuclearTheory}. 

Modern single-particle and {\em ab initio} approaches, while powerful for light nuclei, 
struggle to describe collective motion in heavier systems due to the increasing number
of active degrees of freedom~\cite{Hergert2020AbInitioChallenges, Ekstrom2023AbInitioDefinition}.
As a result, collective models continue to play a crucial role in describing the 
structure of nuclei in regions where deformation dominates. In this work, we focus
on rare-earths (Ce to Yb) and a few heavier neighbors (Hf to Os), where nuclear structure
evolves from near-spherical to highly deformed shapes. Since this region exhibits a broad
range of quadrupole collectivity and includes some of the most deformed nuclei in the
nuclear chart, it provides an excellent testing ground for collective models. Following
Iachello’s proposal for the symmetries E(5) and
X(5)~\cite{PhysRevLett.85.3580, PhysRevLett.87.052502}, which describe the critical points
of the shape phase transitions from spherical to $\gamma$-unstable nuclei and from
spherical to axially deformed nuclei, respectively, numerous interpretations and extensions
have emerged to predict excitation energies and quadrupole transition strengths associated
with collective motion~\cite{RevModPhys.82.2155}. Among these, the Confined $\beta$-Soft
(CBS) rotor model, introduced by Pietralla {\em et al.}~\cite{PhysRevC.70.011304, Reese2011CBS},
provides a simple, yet effective, analytic solution of the Bohr Hamiltonian restricted
to the axially symmetric case ($\gamma \approx 0^\circ$), bridging the gap between the X(5)
critical-point symmetry and the SU(3) limit (rigid rotor).

\subsection*{Motivation}

The CBS model provides predictions for both excitation energies and reduced electric
quadrupole transition probabilities and is therefore well suited for a systematic study
of collective nuclear structure. In the present work, we apply the model to even-even
rare-earth nuclei with
\( 2.904 < R_{4/2} < 3.333 \),
where \(R_{4/2}=E(4^+_1)/E(2^+_1)\),
corresponding to nuclei lying between the X(5) critical-point symmetry and the
rigid-rotor limit. This interval covers an important region of the nuclear chart where
nuclei are well deformed and collective rotational motion dominates, whereas purely
single-particle approaches become less effective. By focusing on this region, the CBS
rotor model allows for a description of the collective behavior, using a relatively
simple framework to study systematic trends across isotopic chains. At the same time,
deviations from the model predictions can point to the presence of additional physical
effects included in a non-purely collective description.

Moreover, a long-standing open problem in this mass region concerns the nature of
excited \(0^+\) states and their interpretation as bandheads of the so-called
\(\beta\)-bands~\cite{Garrett2001, SHARPEYSCHAFER201045c, sharpey2019stiff,  APRAHAMIAN2025104173},
that is, rotational bands built on top of these states. Despite extensive experimental
and theoretical studies exist, the structure and origin of these states still remain
under debate~\cite{APRAHAMIAN2025104173}.

In the above context, the CBS rotor model can be particularly useful, as it allows
the simultaneous calculation of energy levels and quadrupole transition strengths,
not only within the ground-state band, but for the \(\beta\)-band, as well. By applying
the model systematically to rare-earth even--even nuclei, we aim to investigate the
collective character and evolution of the excited \(0^+\) bandheads and the rotational
bands built upon them. In this way, the present study may guide future experimental
efforts toward targeted measurements of B(E2) transition strengths and excited \(0^+\)
states, while also highlighting the limitations of purely collective models and
pointing to missing physics that should be addressed in more refined theoretical 
approaches.

\section{The CBS Model}
\label{sec:The_Model}

\subsection{Theoretical framework}

\begin{figure*}[h!]
    \centering
    \includegraphics[width=0.6\textwidth]{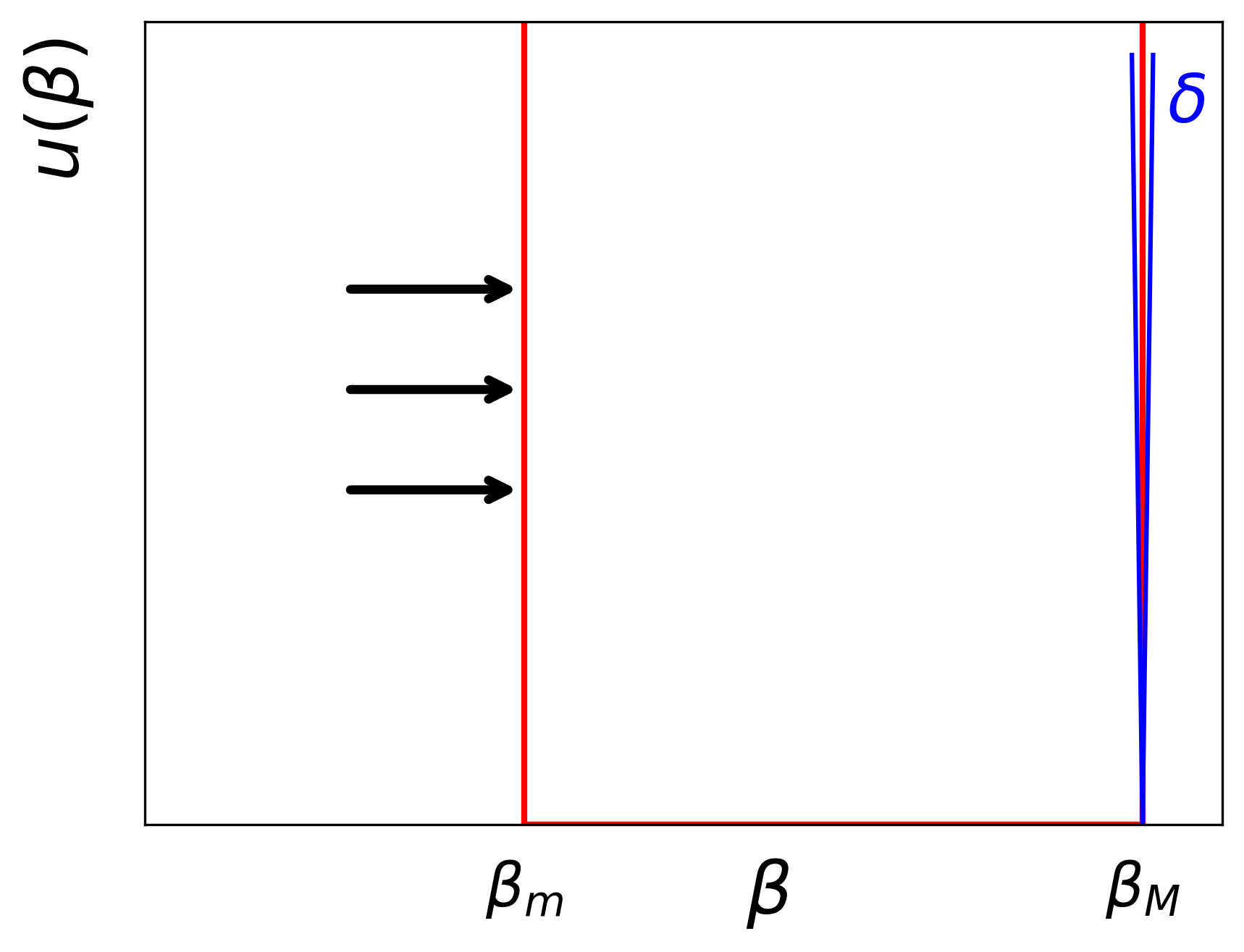}
    \caption{The "moving wall" of the CBS potential, interpolating between the X(5) critical-point symmetry and the rigid-rotor limit.}
    \label{fig:movingwall}
\end{figure*}

We proceed to give a brief description of the CBS model. As already stated, the CBS rotor
model interpolates between the X(5) limit, which corresponds to an analytical solution of
the Bohr Hamiltonian~\cite{PhysRevLett.87.052502}, and the rigid-rotor limit, allowing the
degree of \(\beta\)-softness to vary across this region. The interested reader can refer
to references~\cite{PhysRevC.70.011304, PhysRevC.72.011303} for the full details. The
starting point for this model is the Bohr collective Hamiltonian~\cite{bohr1952coupling,
Bohr1997, Bohr1975NuclearSV}:
\begin{equation}
\hat{H}
=
-\frac{\hbar^{2}}{2B}
\left[
\frac{1}{\beta^{4}}
\frac{\partial}{\partial \beta}
\left(
\beta^{4}
\frac{\partial}{\partial \beta}
\right)
+
\frac{1}{\beta^{2}\sin 3\gamma}
\frac{\partial}{\partial \gamma}
\left(
\sin 3\gamma
\frac{\partial}{\partial \gamma}
\right)
-
\frac{1}{4\beta^{2}}
\sum_{k=1}^{3}
\frac{\hat{Q}_{k}^{2}}
{\sin^{2}\!\left(\gamma - \frac{2\pi k}{3}\right)}
\right]
+
V(\beta,\gamma),
\label{eq:Bohr}
\end{equation}
where $\beta$ and $\gamma$ denote the collective quadrupole deformation
parameters~\cite{bohr1952coupling, Fortunato:2004ij, Bonatsos_2011, Buganu_2016}.
The parameter $\beta$ describes the magnitude of the quadrupole deformation (deviation
from sphericity), while $\gamma$ characterizes the deviation of the nuclear shape from
axial symmetry. Quantity $B$ is the collective mass parameter, and the operators
$\hat{Q}_k$ ($k=1,2,3$) represent the components of the angular momentum in the intrinsic
frame. Eq.~\eqref{eq:Bohr} corresponds to a five-dimensional Schrödinger equation which
cannot be solved analytically unless specific simplifying approximations are
introduced~\cite{PhysRevLett.85.3580}. A major issue in solving the eigenvalue problem
with Eq.~\eqref{eq:Bohr} pertains to its non-separability, which can be overcome by
considering a potential $V(\beta,\gamma)= u(\beta)+v(\gamma)$~\cite{PhysRevLett.87.052502}
for axially symmetric prolate $(\gamma \approx 0^{\circ})$ nuclei. Then, the wave functions
can be separated into
\begin{equation}
\Psi(\beta, \theta_i)=\xi_L(\beta)\mathcal{D}^{L}_{M,K}(\theta_{i}),
\label{separated}
\end{equation}
where $\mathcal{D}^{L}_{M,K}$ are the Wigner functions and $\theta_{i}$ the Euler angles
of the intrinsic system. By focusing on the radial part of the Bohr--Hamiltonian, one gets 
\begin{equation}
- \frac{\hbar^2}{2B}
\left[
\frac{1}{\beta^4}\frac{\partial}{\partial \beta}\Bigl(\beta^4\,\frac{\partial}{\partial \beta}\Bigr)
- \frac{L(L+1)}{3\,\beta^2}
+ u(\beta)
\right]
\xi_L(\beta)
= E\,\xi_L(\beta).
\label{radial}
\end{equation}
At this point, the Confined \(\beta \)-Soft rotor model can be applied. Unlike the X(5)
solution, the CBS framework employs a potential well with a ``moving wall''~\cite{RevModPhys.82.2155}
\begin{equation}
u(\beta) =
\begin{cases}
0, & \text{for } \beta \in [\beta_m,\beta_M], \\
\infty, & \text{otherwise}.
\end{cases}
\label{eq:V1_beta_square_well}
\end{equation} which bridges the region between the X(5) critical point ($\beta_m$=0) and the rigid rotor when $\beta_m$=$\beta_M$ (\ref{fig:movingwall}). 

Inserting this potential in \eqref{radial} leads to a Bessel equation of the form:
\begin{equation}
\tilde{\xi}'' + \frac{\tilde{\xi}'}{z}
+ \left(1 - \frac{\nu^2}{z^2}\right)\tilde{\xi} = 0,
\label{eq:bessel_equation}
\end{equation} where \(z = \sqrt{\frac{E}{\hbar^{2}/2B}}\,\beta
,\quad \tilde{\xi}(z) = \beta^{3/2}\,\xi_L(\beta)\) and $\nu = \sqrt{\dfrac{L(L+1)}{3} + \dfrac{9}{4}}$. Eq. \eqref{eq:bessel_equation} has solutions that are linear combinations of the Bessel functions
$J_{\nu}(z)$ and $Y_{\nu}(z)$ of order $\nu$:
\begin{equation}
\tilde{\xi}_{\nu}(z) \propto A\,J_{\nu}(z) + B\,Y_{\nu}(z),
\label{eq:bessel_solutionconst}
\end{equation} which can also be written as
\begin{equation}
\tilde{\xi}_{\nu}(z) \propto J_{\nu}(z) + \gamma\, Y_{\nu}(z) \, 
\label{eq:bessel_solution2}
\end{equation} where $\gamma = \frac{B}{A}$. Following the derivation in~\cite{PhysRevC.70.011304}, one obtains the energy eigenvalues
\begin{equation}
E_{L,s}
=
\frac{\hbar^{2}}{2B}
\left(
\frac{z^{(r_\beta)}_{L,s}}{\beta_M}
\right)^{2} \,,
\label{eq:energy_spectrum}
\end{equation}
where $B$ denotes the mass parameter of the liquid drop model. An analytical description
of the origin of this parameter can be found in~\cite{Fortunato:2004ij, Buganu_2016}. These eigenvalues depend on a single parameter, $r_{\beta}$, defined as:
\begin{equation}
r_\beta = \frac{\beta_m}{\beta_M},
\label{eq:r_beta}
\end{equation}
used to characterize the relative position of the inner and outer boundaries of the
$\beta$-potential well. It defines the ratio of these boundaries and determines the
stiffness of the $\beta$ degree of freedom~\cite{PhysRevC.70.011304}. It also takes
values in the interval \(0 \le r_{\beta} \le 1\), with the limits \(r_{\beta}=0\) and
\(r_{\beta}=1\) corresponding to the X(5) critical-point symmetry and the rigid-rotor
limit, respectively. For each isotope, $r_{\beta}$ can be determined by fitting the
calculated ground-state energy ratios to the experimental ones. In practice, a trial
value of $r_{\beta}$ fixes the corresponding Bessel zeros $z_{L,1}(r_{\beta})$, from 
which the normalized excitation energies are computed. The value of $r_{\beta}$ is
varied until the deviation between calculated and experimental ratios is minimized.
Smaller $r_{\beta}$ values produce softer, transitional behavior near the X(5) limit
($R_{4/2} = 2.904$), whereas larger $r_{\beta}$ correspond to a stiffer $\beta$
potential approaching the rigid-rotor limit ($R_{4/2} = 3.333$), associated with
the SU(3) limit~\cite{Papadopoulos2026}. The CBS model was successfully used in the
past to describe the even-even \isotope[152-156]{Sm}~\cite{PhysRevC.70.011304} and
\isotope[150-156]{Nd}~\cite{PhysRevC.72.011303} isotopes analytically, showing the
capabilities of the model. Moreover, regarding the calculation of the B(E2)
transition strengths, the following E2 operator is employed:
\begin{equation}
T_{\mu}^{\Delta K=0}(E2)
=
e^{(1)}\,\beta_M \,\langle \cos\gamma \rangle_{\gamma}
\left(\frac{\beta}{\beta_M}\right)
-
\sqrt{\frac{2}{7}}\,
e^{(2)}\,\beta_M
\left[
\frac{\langle \cos 2\gamma \rangle_{\gamma}}{\langle \cos\gamma \rangle_{\gamma}}
\left(\frac{\beta}{\beta_M}\right)^2
\right]
D^{2}_{\mu 0}(\Omega)\,,
\label{eq:T_E2_DeltaK0_full}
\end{equation}
which can be written as
\begin{equation}
T_{\mu}^{\Delta K=0}(E2)
=
e_{\mathrm{eff}}\left[
\left(\frac{\beta}{\beta_M}\right)
+
\chi\left(\frac{\beta}{\beta_M}\right)^2
\right]
D^{2}_{\mu 0}(\Omega)\,,
\label{eq:T_E2_DeltaK0_eff}
\end{equation}
with the effective charge
\begin{equation}
e_{\mathrm{eff}} = e^{(1)}\,\beta_M \langle \cos\gamma \rangle_{\gamma}
\label{eq:effcharge}
\end{equation}
and
\begin{equation}
\chi = -\sqrt{\frac{2}{7}}\,e^{(2)}\,\beta_M
\frac{\langle \cos 2\gamma \rangle_{\gamma}}{\langle \cos\gamma \rangle_{\gamma}}
\label{eq:chi}
\end{equation}
to be adjusted to available experimental B(E2) transition data~\cite{PhysRevC.70.011304}.
Here, \(e^{(1)}\) and \(e^{(2)}\) denote the effective quadrupole charge parameters
associated with the linear and quadratic terms of the collective \(E2\) transition
operator, respectively. Overall, it can be seen that the CBS rotor model uses only two
dimensionless parameters, $r_\beta$ and $\chi$, and two scale factors,
$\hbar^2/(2B\beta_M^2)$ and $e_{\mathrm{eff}}$.

\subsection{Methodology}

This section describes the practical computational procedure followed to obtain numerical
results within the CBS rotor model. The emphasis is placed on how the model is implemented
and used in practice, rather than on its theoretical formulation, which has been fully
described in the original publication~\cite{PhysRevC.70.011304}. The model starts by fitting
on experimental excitation energies of the ground state band, at least up to the $6^+_1$ state.
In this work, we feed experimental excitation energies up to the $8^+_1$ or, where available,
the $10^+_1$ state, to achieve the best possible determination of the $r_{\beta}$ parameter.
From this fit, the values of the parameter $r_{\beta}$ and the energy scale factor
${B\,\beta_{M}^2}$ can be obtained. The parameter $r_\beta$ is determined by minimizing the weighted $\chi^2$ (not to be confused with the parameter $\chi$ in Eq.~\eqref{eq:chi}):
\begin{equation}
    \chi^2 = \sum_{I=2}^{I_\mathrm{max}} \frac{\left(E(I_1^+)_\mathrm{exp} - E(I_1^+; CBS)\right)^2}{\sigma^2_{I,\mathrm{exp}}},
\end{equation}
where the sum runs over the measured ground-state band energies from $I=2$ up to $I_\mathrm{max}$ (8 or 10, depending on data availability), and $\sigma_{I,\mathrm{exp}}$ is the experimental uncertainty on each level energy. This is handled internally by the CBS code~\cite{Reese2011CBS}. 

Regarding the calculation of B(E2) transition strengths,
we use the experimental values of
${B(E2;2_1^+ \rightarrow 0_1^+)}$ and
${B(E2;4_1^+ \rightarrow 2_1^+)}$, where available, to determine the value of ${\beta_{M}}$.
At this point it should be stressed that, as stated in~\cite{Kocheva2026Er}, while the CBS
code enables either a constrained fit of ${\beta_M}$ with fixed $r_\beta$ and ${B\beta_{M}^2}$,
or a simultaneous fit of all three parameters, both cases result in identical parameter values.
We therefore choose to simultaneously fit all the parameters, as well as the $\chi$ parameter
and the scale factor $e_{\mathrm{eff}}$ in Eq.~\eqref{eq:T_E2_DeltaK0_eff}. Especially for
the calculation of the energies, the following formula is employed~\cite{Reese2011CBS}:
\begin{equation}
E(L,s)=
\begin{cases}
E_0 + \dfrac{z_{L,s}^2 - z_{0,s}^2}{2B\,\beta_{\max}^2}
+ \epsilon\,L(L+1),
& s = 1, \\[8pt]
b_s \bigl(E(L,s) + b_s \bigr),
& \text{else}.
\end{cases}
\end{equation}

\section{Results and Discussion}
\label{sec:comp}

 In this section we proceed to list the literature sources from which the experimental
 data were obtained (Sec.~\ref{sec:data}), and give some general remarks regarding
 theoretical predictions (Sec.~\ref{sec:remmarks}). Following that, a short discussion
 is presented for each individual isotopic chain under consideration
 (Sec.~\ref{sec:Ce}-\ref{sec:Os})

\subsection{Experimental data sources}
\label{sec:data}

Experimental excitation energies for the ground-state band were retrieved from the ENSDF
(NUDAT) database~\cite{nudat} and from evaluated data compiled in
Nuclear Data Sheets~\cite{Basu, martin_2013, NICA20252, IIMURA20221, ELEKES2015191, REICH20122537, NICA20171, Nica:2021cny, Nica:2024ttv, singh_chen_2018, Baglin2008, Baglin2010, BAGLIN20181, Singh1995, basunia_2006, Achterberg2009, McCutchan2015, SINGH201521, BAGLIN2010275, BATCHELDER20221, KONDEV20181, Singh:2020zfc}.
For the comparison of the model results for the reduced electric quadrupole transition
probabilities B(E2), we used the recommended values from the tables of E2 transition
probabilities~\cite{PRITYCHENKO20161}, supplemented by evaluated data from
Nuclear Data Sheets~\cite{Basu, martin_2013, Reich:2009cea, REICH20122537, NICA20171, Nica:2021cny, Nica:2024ttv, singh_chen_2018, Baglin2008, Baglin2010, BAGLIN20181, Singh1995, basunia_2006, Achterberg2009, McCutchan2015, SINGH201521, BAGLIN2010275, BATCHELDER20221, KONDEV20181, Singh:2020zfc}.
In addition, a dedicated literature search was performed to identify recently published,
non-evaluated experimental data~\cite{Kocheva2026Er, PhysRevC.101.024313, PETKOV2017240, PhysRevC.95.034316, PhysRevC.91.044301, PhysRevC.99.024316, PhysRevC.106.024326, PhysRevC.108.024305}, which were also included in the present investigation.

\subsection{General remarks}
\label{sec:remmarks}

\begin{figure*}[ht]
    \centering
    \includegraphics[width=0.97\textwidth]{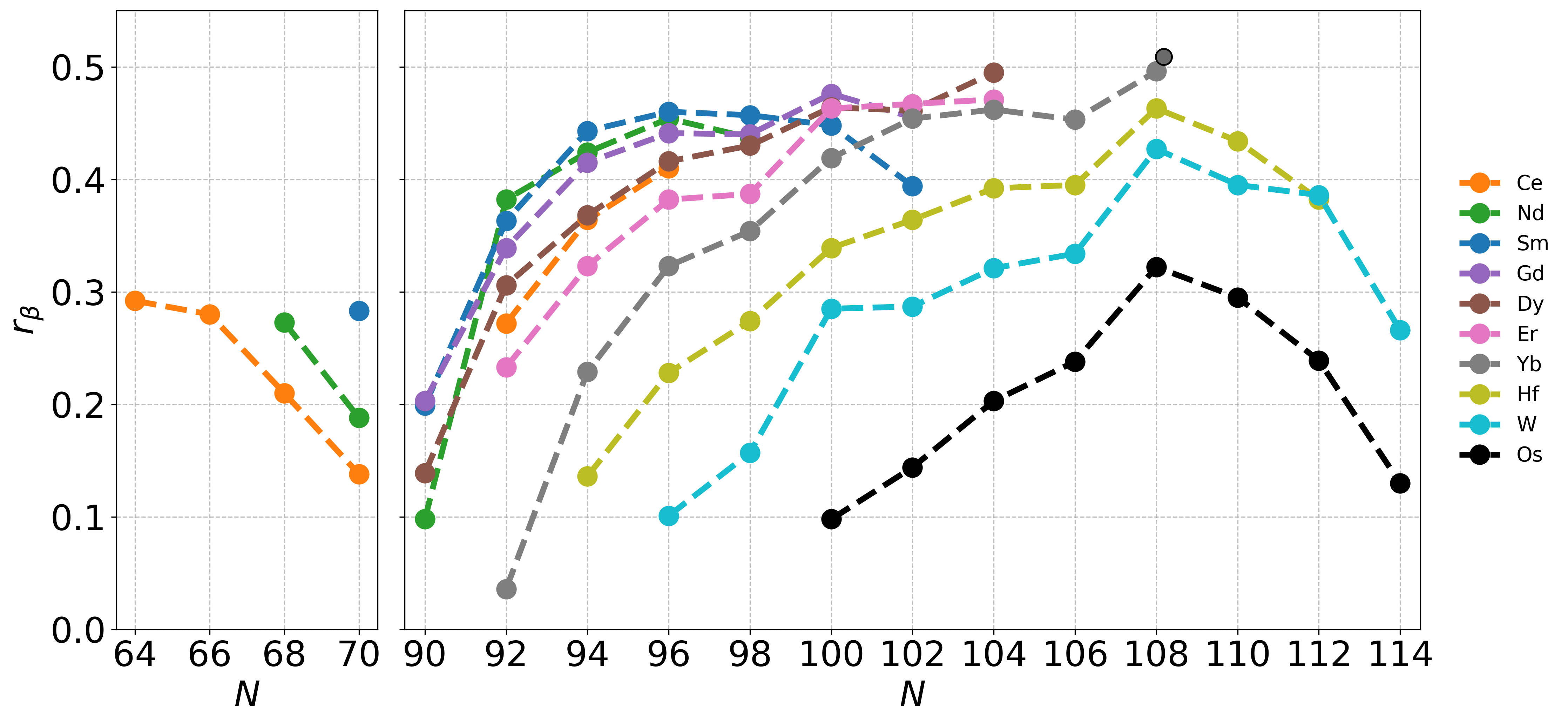}
  \caption{
Evolution of the CBS parameter $r_\beta$ along the even--even isotopic chains of Ce-Os elements as a function of neutron number N. The bold gray point at \isotope[178]{Yb} ($N=108$) represents the value $r_\beta$ = 0.509, obtained from the CBS calculation, based on new measurements reported in \cite{koseoglou_inprep}.}
    \label{fig:rb}
\end{figure*}

The predictions for the excited energy states and B(E2) transition ratios of the
ground-state band of even-even rare-earth nuclei up to the $10^+$, together with the
corresponding experimental values, are presented in Tables~\ref{tab:energy_ratios}
and~\ref{tab:be2_ratios}, respectively. Table~\ref{tab:beta_band_energies_cbs} summarizes
the excitation energies of the $\beta$-band up to the $6^+$ state in keV, along with
the corresponding quadrupole transition-strength ratios. Fig.~\ref{fig:rb} provides
a systematic overview of the extracted $r_\beta$ values for each isotope studied in
the present work.

As seen in Table~\ref{tab:energy_ratios}, the CBS rotor model reproduces the ground-state
band energies of even–even nuclei in the rare-earth region with high precision, showing
the capability of the model to find a $r_\beta$ value to fit well each of the isotopes
examined in the region. This level of agreement is consistent with the interpretation
that even–even rare-earth nuclei exhibit strong quadrupole collectivity and relatively
stable axial deformation, especially in their low-lying states. In this mass region, the
low-lying excitation spectrum is dominated by rotational motion of a deformed intrinsic
shape, and collectivity reduces the influence of single-particle degrees of
freedom~\cite{Bohr1975NuclearSV}. The CBS results presented in Table~\ref{tab:be2_ratios}
show that the model reproduces the overall trend of the B(E2) transition ratios for the
lowest excited states reasonably well, within the experimental uncertainties. In contrast,
in many cases, the model fails to reproduce the observed systematics of these transition
ratios at higher spins. Since this discrepancy cannot be attributed to a simple overall
scaling, it suggests a deeper physical origin that warrants further investigation.

In particular, CBS predictions generally show smooth increasing ratios with angular
momentum \(L\), revealing a higher rigidity. The increase in the average deformation
$\left\langle\beta\right\rangle_L$ with angular momentum is referred to as centrifugal
stretching and is caused by the \(L(L+1)\) centrifugal term in
Eq.~\eqref{radial}~\cite{PhysRevC.72.011303, PhysRevC.79.024307}.
However, as shown in Table~\ref{tab:be2_ratios}, the experimental values do not always
follow this monotonic trend. For example, some nuclei exhibit
$R_{6/2}<R_{4/2}$,
$R_{10/2}<R_{8/2}$,
or irregular jumps, such as:
$
R_{4/2} = 1.559 \rightarrow
R_{6/2} = 1.418 \rightarrow
R_{8/2} = 1.856 \rightarrow
R_{10/2} = 1.336
$
(see \isotope[162]{Yb}).
In many cases, experimental values for $R_{10/2}$ drop after peaking around $R_{6/2}$
or $R_{8/2}$. Such behavior is commonly associated with the onset of backbending,
a well-known phenomenon in deformed nuclei whereby the moment of inertia increases
abruptly with spin~\cite{STEPHENS1972257, Wyss03042022}. Microscopically, backbending
is interpreted as the alignment of high-j quasiparticles --typically pairs of neutrons
or protons-- along the rotational axis~\cite{book}. This reflects the model assumption
of persistent rigidity at higher angular momentum, which is not always valid in real
nuclei. As a result, CBS predictions can overestimate the degree of rigidity, especially
for large values of $r_\beta$, which correspond to stronger confinement in the $\beta$-degree 
of freedom. In such cases, the model tends to reproduce rigid-rotor–like behavior, while
experimental data often reveal saturation or even a reduction in rigidity with increasing
spin. Moreover, irregular changes in the experimental $B(E2)$ ratios may suggest other
phenomena not captured by the CBS framework, such as band crossings, configuration changes
(e.g., alignment of an $i_{13/2}$ neutron orbital) and mixing with non-yrast or
two-quasiparticle states~\cite{StephensSimon1972, Stephens1975, BengtssonFrauendorf1979}.
Since CBS is a purely collective model based on the Bohr Hamiltonian, it does not incorporate
such microscopic effects~\cite{PhysRevC.72.011303}. While it provides useful insight into
collective trends, its predictive power at high angular momentum is inherently limited.
It should also be noted that the $\gamma$ degree of freedom is not explicitly taken into
account in the CBS model. In this framework, the potential $V(\beta,\gamma)$ in
Eq.~\eqref{eq:Bohr} is minimized at $\gamma \approx 0^\circ$, corresponding to small
harmonic oscillations around axial symmetry. As a result, the model is effectively restricted
to axially symmetric shapes and does not describe $\gamma$-softness. However, several works
have shown that the $\gamma$ degree of freedom can play an important role in the same regions
of the nuclear chart (see, e.g., \cite{Koseoglou_Werner_Pietralla_Bonatsos_2019, PhysRevC.101.014303, koseoglou_werner_pietralla_2022}).
This limitation should therefore be kept in mind when interpreting the results presented
in the plots and tables below.

Another important observation in Table~\ref{tab:be2_ratios} is the scarcity of experimental
data for quadrupole transitions, particularly between higher-spin states. This underscores
the need for new and precise measurements to better constrain theoretical models and improve
our understanding of nuclear structure.

Regarding the CBS calculations of the excited $0_\beta^+$ states in Table~\ref{tab:beta_band_energies_cbs}, we observe the following correlation between the
$r_{\beta}$ parameter and the evolution of these states: for smaller values of $r_{\beta}$
close to the X(5) limit, the potential in $\beta$ is softer and thus the \(\beta\)-band
appears at lower excitation energy, while in nuclei close to the rigid-rotor limit, where
\(r_{\beta}\) approaches 1, the potential becomes narrow and stiff and this strong
confinement of the \(\beta\)-motion results in a large restoring force, rapidly pushing
the \(\beta\)-bandhead to higher excitation energies~\cite{PhysRevC.70.011304}. At the same
time, Table~\ref{tab:beta_band_energies_cbs} provides a useful set of predictions for the
$\beta_1$ band, which may assist in its identification among several experimentally
observed excited $K=0^+$ bands in a given nucleus. Such identification is often nontrivial,
as the nature of excited $0^+$ states remains an open question: they may correspond to
vibrational excitations about the deformed ground-state shape, to bandheads associated
with coexisting shape minima, or to quasiparticle excitations~\cite{APRAHAMIAN2025104173}.
Therefore, the present results may serve as a practical tool for the interpretation of
experimental data and guidance for future studies.

Before discussing each isotopic chain separately, Fig.~\ref{fig:rb} provides an overall
view of the $r_\beta$ values for all the rare-earth nuclei examined in the present work.
The parameter $r_\beta$ has been shown to correlate with the ground-state band ratio
$R_{4/2}$~\cite{PhysRevC.72.011303, PhysRevC.70.011304}, a trend that is also consistent
with the evolution of the energy ratios displayed in Figs.~\ref{fig:ce_energy}--\ref{fig:os_energy}.
An interesting feature of Fig.~\ref{fig:rb} is the systematic evolution of the $r_\beta$
values across the rare-earth region. The largest value is found for $^{178}\mathrm{Yb}$,
with $r_\beta = 0.509$~\cite{koseoglou_inprep}, identifying it as the most rigidly deformed
nucleus in the present set. More generally, the figure reveals a clear structural trend:
as the proton number $Z$ increases, the maximum of $r_\beta$, and hence the peak of
rotational rigidity, shifts toward larger neutron number $N$. This shift of the maximum
towards higher neutron number with increasing proton number reflects the need to maintain
optimal quadrupole correlations~\cite{Casten1985}.

\subsection{Ce (Z=58)}
\label{sec:Ce}

\begin{figure*}[ht]
    \centering
    \includegraphics[width=0.98\textwidth]{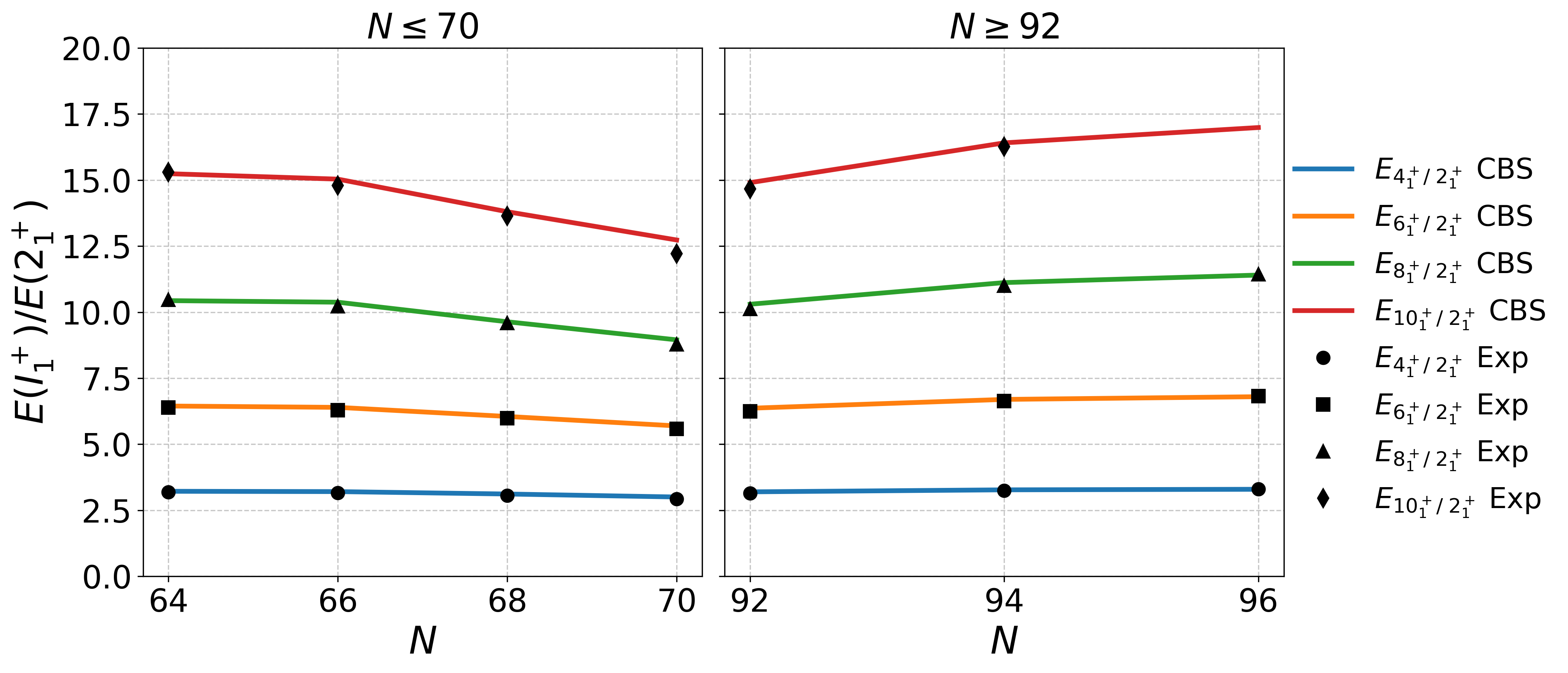}
    \caption{
Systematics of energy ratios
$E(I_1^+)/E(2_1^+)$
for the ground-state band of even--even Ce (Z=58) isotopes as a function of neutron number $N$.
Solid lines indicate CBS predictions, while experimental values are shown as symbols. Experimental uncertainties are smaller than the symbol sizes. 
}
    \label{fig:ce_energy}
\end{figure*}

\begin{figure*}[ht]
    \centering
    \includegraphics[width=0.7\textwidth]{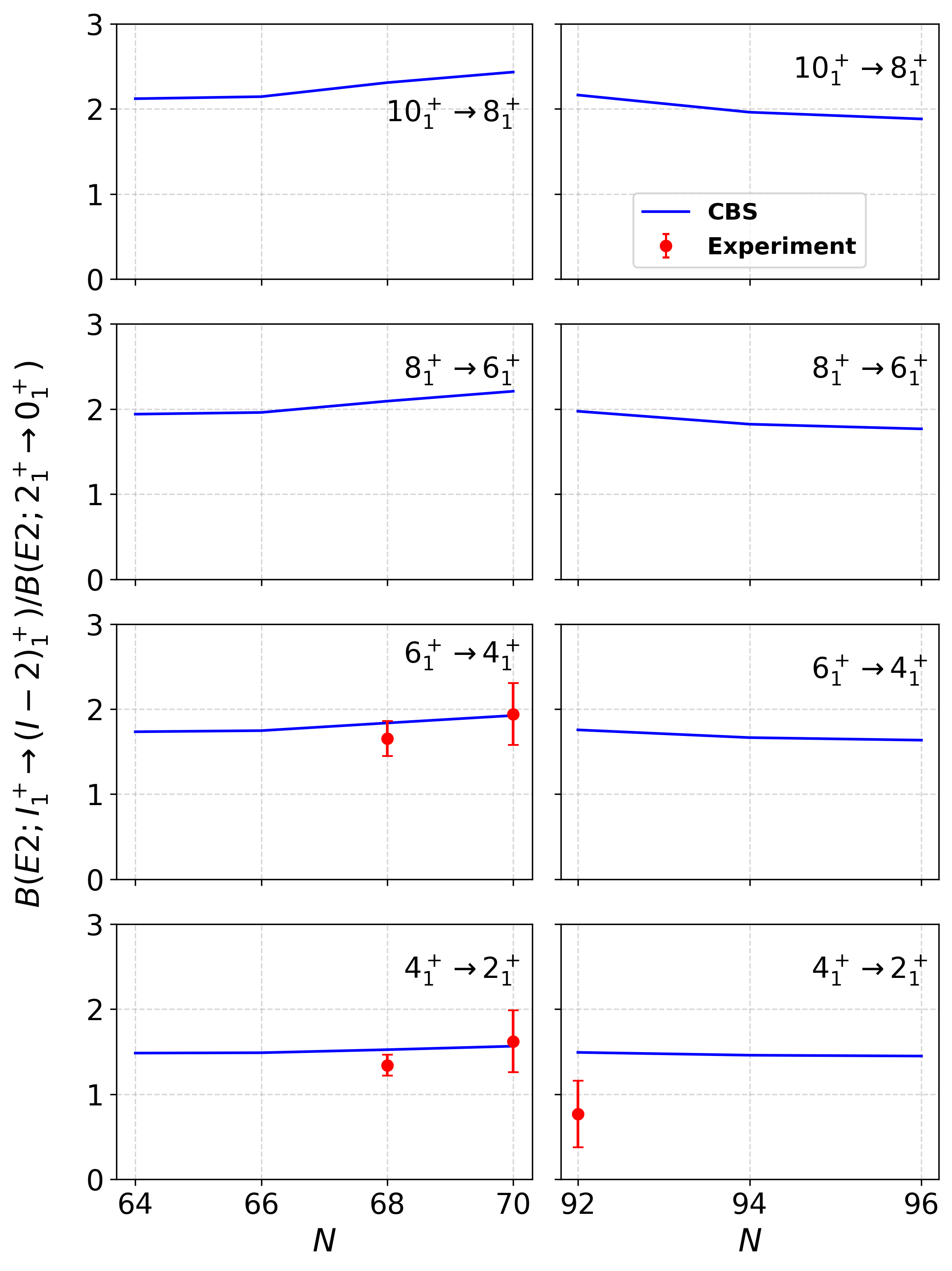}
    \caption{Systematics of reduced electric quadrupole transition ratios
$B(E2; I_1^+ \rightarrow I_1^+-2) / B(E2; 2_1^+ \rightarrow 0_1^+)$
for the ground-state band of even--even Ce (Z=58) isotopes as a function of neutron number $N$. Solid lines indicate CBS predictions, while experimental values are shown as symbols. Experimental data were retrieved from~\cite{IIMURA20221, ELEKES2015191, Basu}. When multiple measurements were available, the most recent values were used.}
    \label{fig:ce_BE2}
\end{figure*}

For the Ce isotopic chain, Fig.~\ref{fig:ce_energy} shows the evolution of the energy ratios 
$E(I_1^+)/E(2_1^+)$ for Ce isotopes as a function of neutron number, separated into two
regions ($N \leq 70$ and $N \geq 92$). The CBS calculations reproduce the overall experimental
trends in both regions. In the lighter isotopes ($N \leq 70$), a gradual decrease of the
ratios with increasing neutron number is observed, indicating a reduction in rotational
rigidity, which is consistently captured by the model. In contrast, for the heavier isotopes
($N \geq 92$), the ratios increase with neutron number, reflecting a transition toward more
rigid rotational behavior. It should also be noted that more neutron-rich isotopes, such as
\isotope[156]{Ce} and \isotope[158]{Ce}, may also fall within the interval
$2.904 < R_{4/2} < 3.333$, although no experimental data are currently available.
Fig.~\ref{fig:ce_BE2} mainly highlights the scarcity of experimental results on quadrupole
transition strengths in the Ce isotopic chain.

\subsection{Nd (Z=60)}
\label{sec:Nd}

\begin{figure*}[ht]
    \centering
    \includegraphics[width=0.98\textwidth]{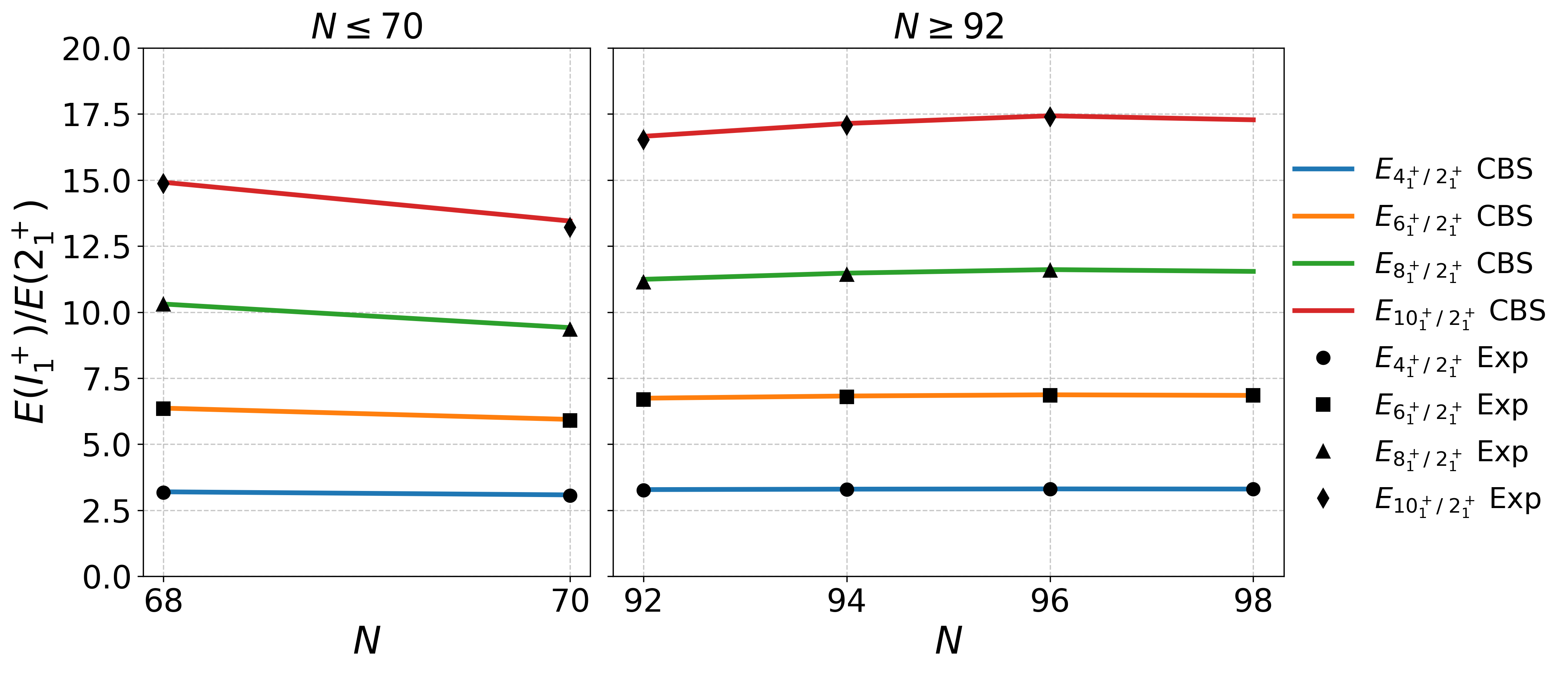}
    \caption{
Systematics of energy ratios
$E(I_1^+)/E(2_1^+)$
for the ground-state band of even--even Nd (Z=60) isotopes as a function of neutron number $N$.
Solid lines indicate CBS predictions, while experimental values are shown as symbols. Experimental uncertainties are smaller than the symbol sizes. 
}
    \label{fig:nd_energy}
\end{figure*}

\begin{figure*}[ht]
    \centering
    \includegraphics[width=0.725\textwidth]{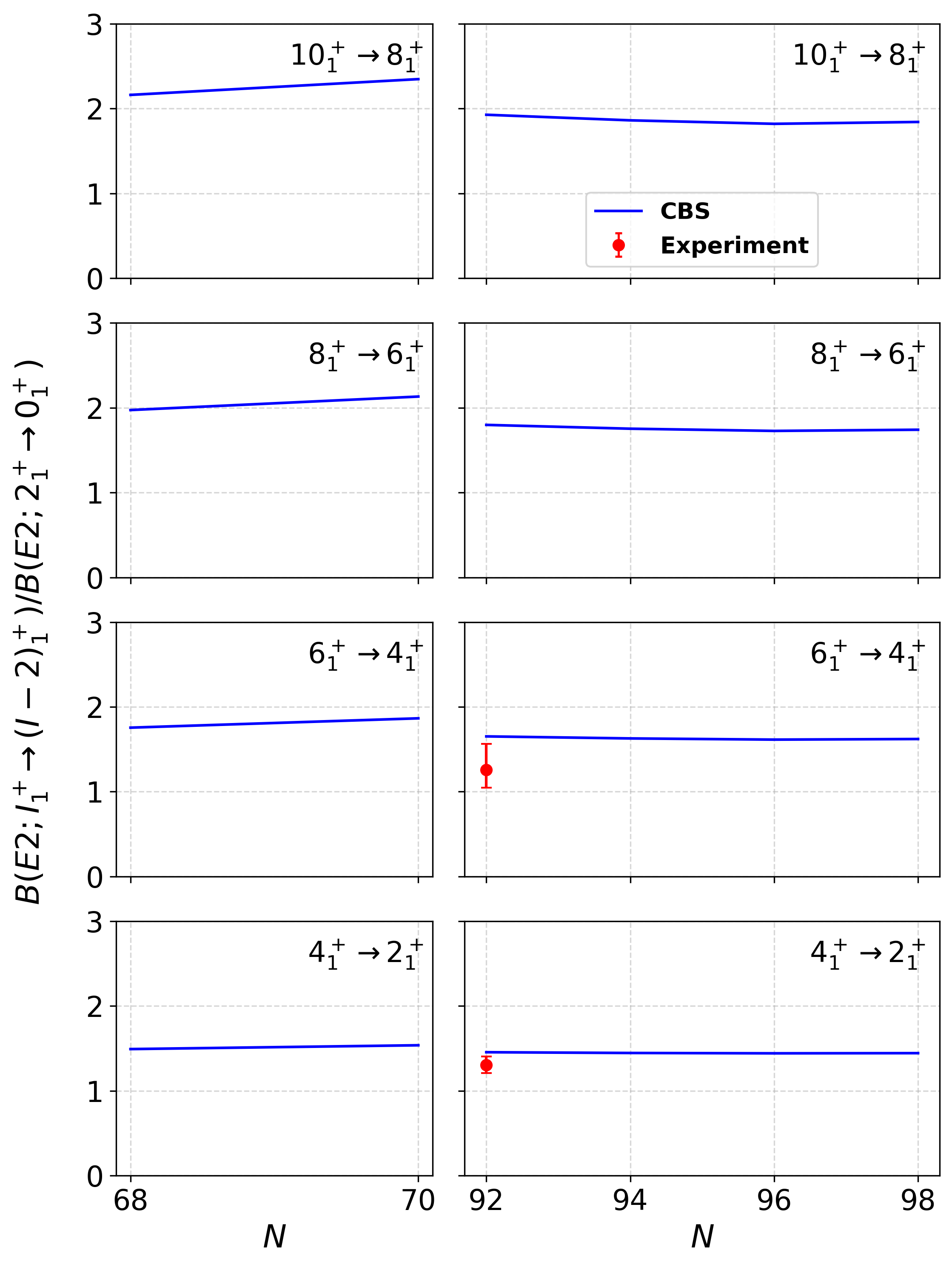}
    \caption{Systematics of reduced electric quadrupole transition ratios
$B(E2; I_1^+ \rightarrow I_1^+-2) / B(E2; 2_1^+ \rightarrow 0_1^+)$
for the ground-state band of even--even Nd (Z=60) isotopes as a function of neutron number $N$. Solid lines indicate CBS predictions, while experimental values are shown as symbols. Experimental data were retrieved from~\cite{IIMURA20221, martin_2013}. When multiple measurements were available, the most recent values were used.}
    \label{fig:nd_BE2}
\end{figure*}

Similarly to the case of Ce isotopes, the experimental data for B(E2) transitions in Nd
isotopes are rather limited, as seen in Fig.~\ref{fig:nd_BE2}. Nevertheless, a noticeable
decrease in the $B(E2)$ ratios is observed between \isotope[150]{Nd} ($N=90$) and
\isotope[152]{Nd} ($N=92$). Within the CBS framework, this behavior is correlated with
the significant change in the $r_\beta$ parameter, which increases from $0.098$ for
\isotope[150]{Nd} to $0.382$ for \isotope[152]{Nd}. The higher $r_\beta$ value for
\isotope[152]{Nd} indicates reduced $\beta$-softness and a tendency toward a more
stabilized rotational structure, resulting in lower transition-strength ratios. This
structural difference is also reflected in the energy systematics shown in
Fig.~\ref{fig:nd_energy}, particularly for the higher-lying $8^+_1$ and $10^+_1$ states,
where centrifugal stretching effects become more pronounced. It should also be noted
that no experimental data are currently available for the $8^+_1$ and $10^+_1$ excited
states of \isotope[158]{Nd} in Fig.~\ref{fig:nd_energy}. Therefore, only the experimental
energies up to $6^+_1$ were used in the fit.

\subsection{Sm (Z=62)}
\label{sec:Sm}

\begin{figure*}[ht]
    \centering
    \includegraphics[width=0.75\textwidth]{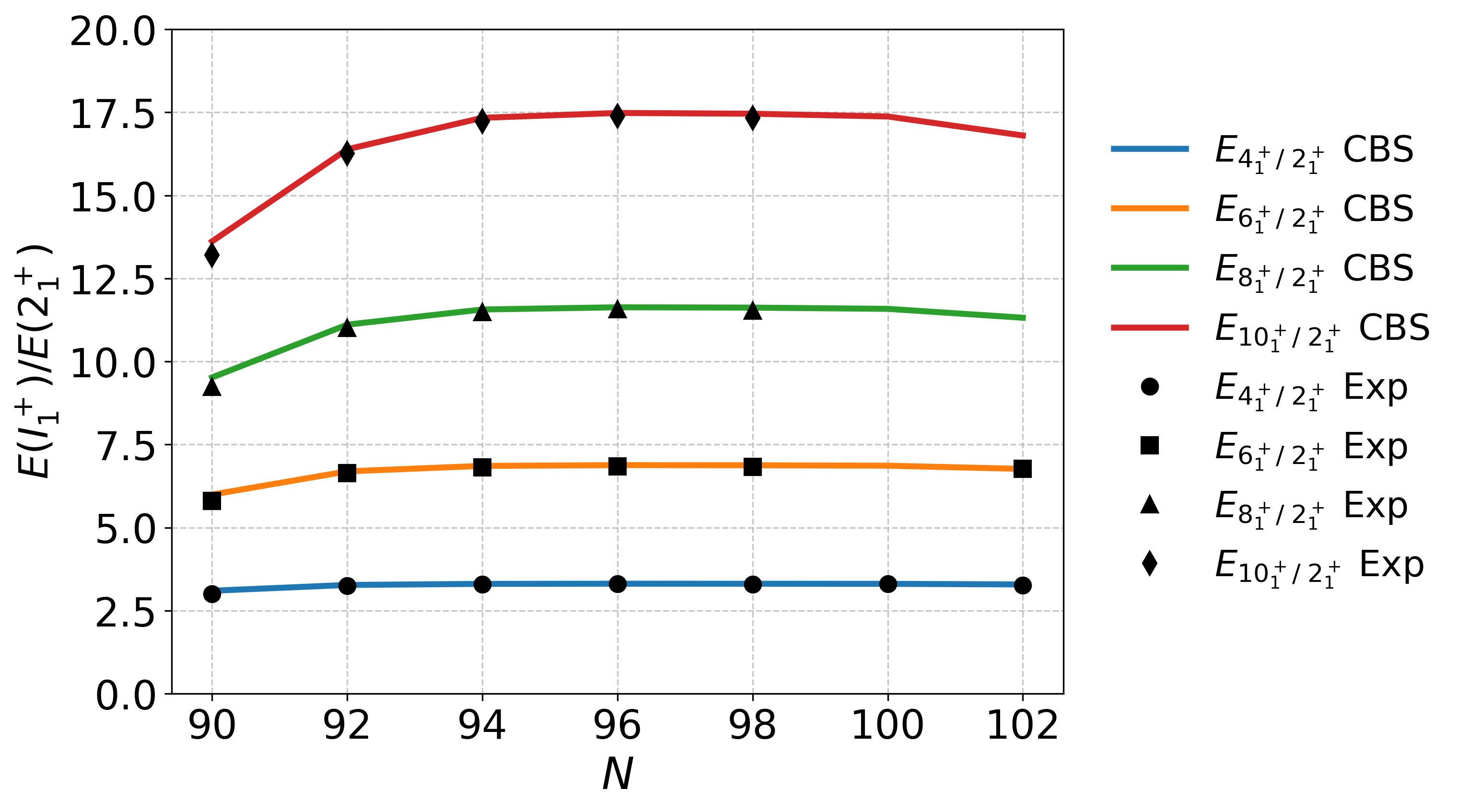}
    \caption{
Systematics of energy ratios
$E(I_1^+)/E(2_1^+)$
for the ground-state band of even--even Sm (Z=62) isotopes as a function of neutron number $N$. Solid lines indicate CBS predictions, while experimental values are shown as symbols. Experimental uncertainties are smaller than the symbol sizes. The CBS results for \isotope[132]{Sm} are omitted from the figure for clarity and can be found in Tables~\ref{tab:energy_ratios}, \ref{tab:be2_ratios}, and \ref{tab:beta_band_energies_cbs}.
}
    \label{fig:sm_energy}
\end{figure*}

\begin{figure*}[ht]
    \centering
    \includegraphics[width=0.7\textwidth]{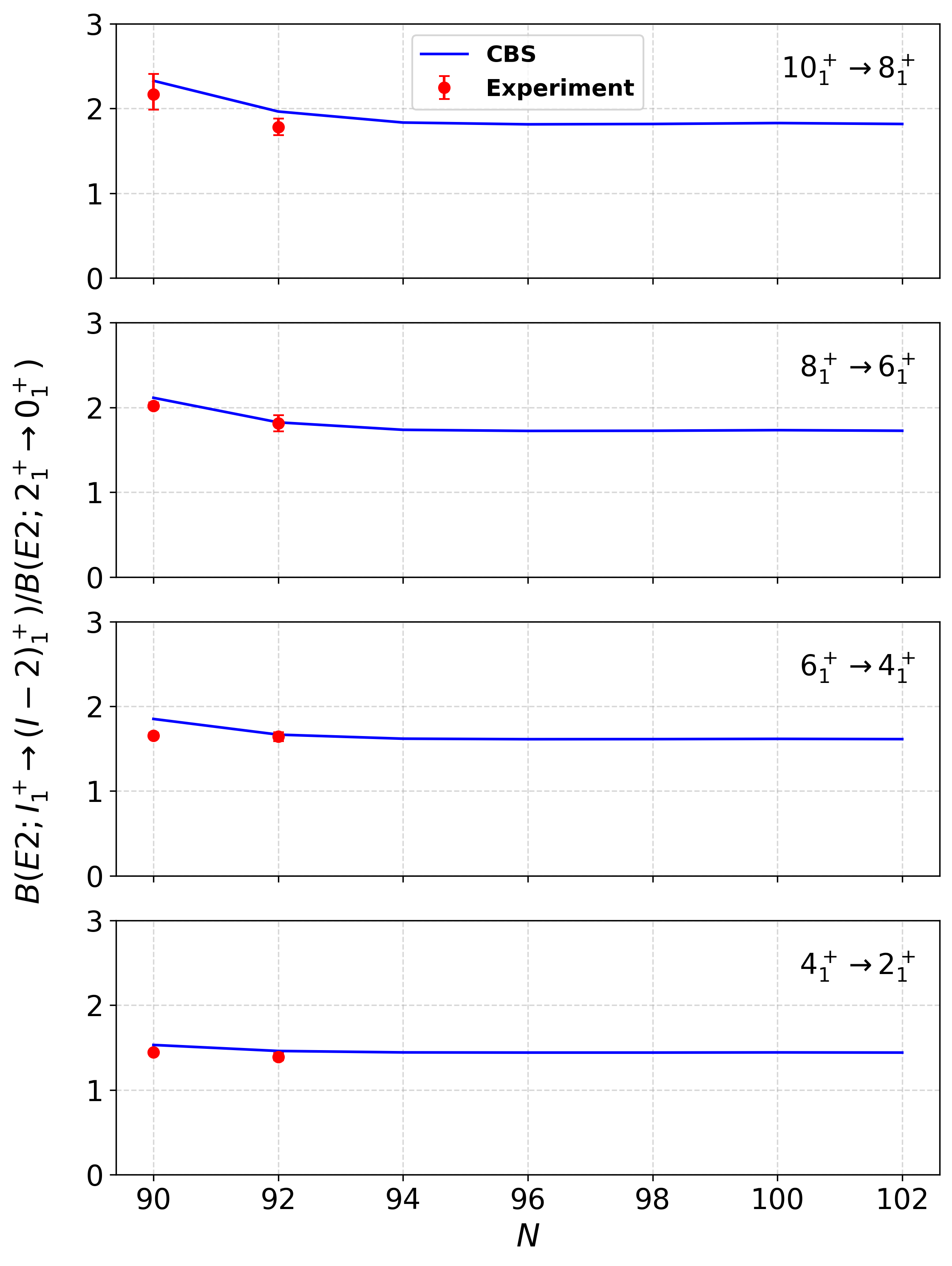}
    \caption{Systematics of reduced electric quadrupole transition ratios
$B(E2; I_1^+ \rightarrow I_1^+-2) / B(E2; 2_1^+ \rightarrow 0_1^+)$
for the ground-state band of even--even Sm (Z=62) isotopes as a function of neutron number $N$. Solid lines indicate CBS predictions, while experimental values are shown as symbols. Experimental data were retrieved from~\cite{martin_2013, NICA20252}. When multiple measurements were available, the most recent values were used. The CBS results for \isotope[132]{Sm} are omitted from the figure for clarity and can be found in Tables~\ref{tab:energy_ratios}, \ref{tab:be2_ratios}, and \ref{tab:beta_band_energies_cbs}.}
    \label{fig:sm_BE2}
\end{figure*}

For Sm isotopes, the saturation of the energy ratios in Fig.~\ref{fig:sm_energy} becomes
evident beyond $N=104$~\cite{FENG1988156}. An important case is that of \isotope[162]{Sm}
($N=100$), for which no experimental data are available for the $6^+_1$ state of the
ground-state band; consequently, a direct CBS fit could not be performed. To complete the
isotopic chain, we estimated the energy of the $6^+_1$ state. As a first step, we
interpolated the $6^+_1$ energy of \isotope[162]{Sm} by averaging the $R_{6/2}$ ratios
of the neighboring isotopes \isotope[160]{Sm} and \isotope[164]{Sm}. The interpolated
ratio for \isotope[162]{Sm} is $R_{6/2}=(6.830+6.770)/2=6.800$. Multiplying by the
experimental value of $E_{1}^{2^+}$ for \isotope[162]{Sm}, we obtain  $E_1^{6^+}=485.5$~keV. However, this approach did not lead to a satisfactory CBS fit, which is rather expected
since a simple interpolation does not fully account for the structural evolution with
neutron number. A more consistent approach is to use the $R_{6/2}$ ratio of a nucleus
whose $R_{4/2}$ value is as close as possible to that of \isotope[162]{Sm} ($R_{4/2}=3.304$).
From Table~\ref{tab:energy_ratios}, \isotope[172]{Yb} has $R_{4/2}=3.305$, making it an
excellent reference case. Using the corresponding $R_{6/2}$ ratio, we obtain
$E_{1}^{6^+}=489.6$~keV, which leads to a successful CBS fit. In Fig.~\ref{fig:sm_BE2},
experimental data are available only for \isotope[152-154]{Sm}, where a relative agreement
between the calculated and measured values is observed. For the remaining nuclei, we provide
theoretical predictions within the CBS framework. Another interesting case is that of
\isotope[134]{Sm}. While it satisfies the condition \(2.904 < R_{4/2} < 3.333\),
the experimental data available in literature do not suffice for a successful fit by the
CBS, suggesting the experimental data for this nucleus are worth revisiting in the near future.


\subsection{Gd (Z=64)}
\label{sec:Gd}

\begin{figure*}[ht]
    \centering
    \includegraphics[width=0.75\textwidth]{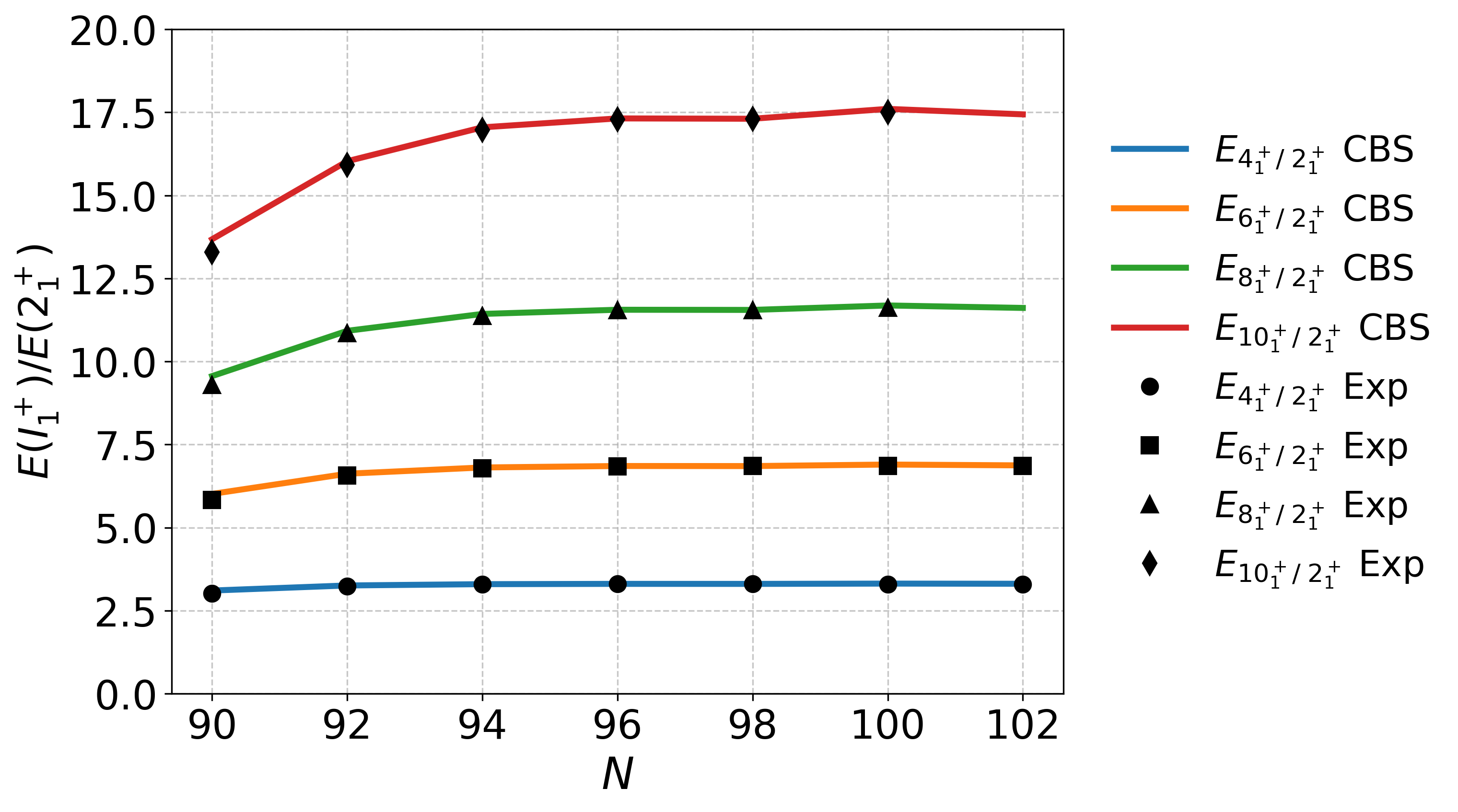}
    \caption{
Systematics of energy ratios
$E(I_1^+)/E(2_1^+)$
for the ground-state band of even--even Gd (Z=64) isotopes as a function of neutron number $N$. Solid lines indicate CBS predictions, while experimental values are shown as symbols. Experimental uncertainties are smaller than the symbol sizes. 
}
    \label{fig:gd_energy_ratios}
\end{figure*}

\begin{figure*}[ht]
    \centering
    \includegraphics[width=0.7\textwidth]{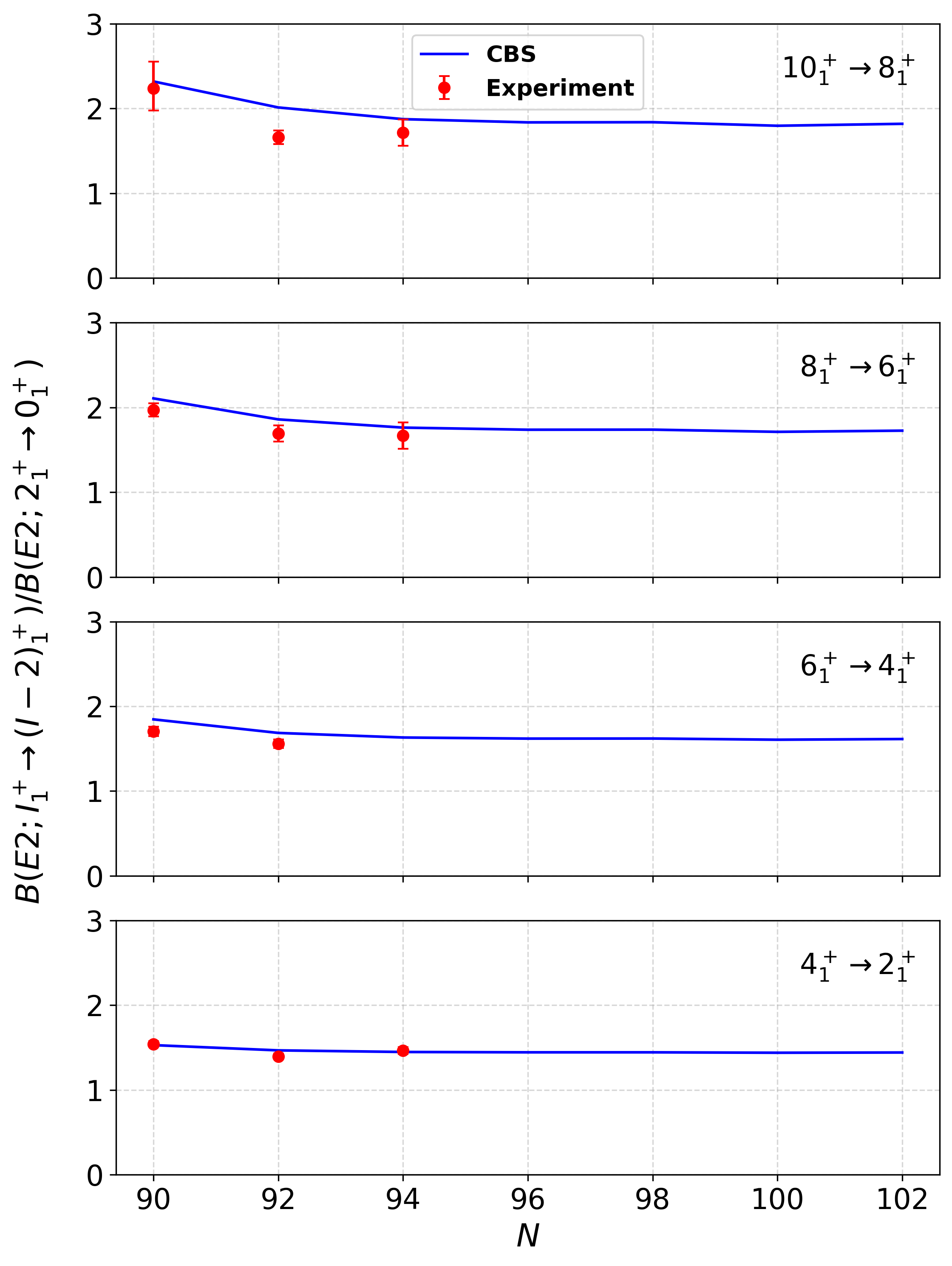}
    \caption{Systematics of reduced electric quadrupole transition ratios
$B(E2; I_1^+ \rightarrow I_1^+-2) / B(E2; 2_1^+ \rightarrow 0_1^+)$
for the ground-state band of even--even Gd (Z=64) isotopes as a function of neutron number $N$. Solid lines indicate CBS predictions, while experimental values are shown as symbols. Experimental data were retrieved from~\cite{NICA20252, REICH20122537, NICA20171}. When multiple measurements were available, the most recent values were used.}
    \label{fig:gd_BE2}
\end{figure*}

Fig.~\ref{fig:gd_BE2} shows that \isotope[154]{Gd} ($N=90$) and \isotope[158]{Gd} ($N=94$)
are well described by the CBS rotor model, as their \(B(E2)\) ratios follow the smooth
increasing trend characteristic of collective rotational behavior. In
Fig.~\ref{fig:gd_energy_ratios}, the CBS calculations reproduce the trend of the experimental
energy ratios in the ground-state band, while also predicting values for the $8^+_1$ and
$10^+_1$ excited states of \isotope[164,166]{Gd}, for which no experimental data are
currently available. A comparison with the available experimental data can also be made
for \isotope[156]{Gd} ($N=92$), where the CBS predictions lie close to the measured values,
except for the \(B_{10/2}\) ratio, for which the model underestimates the transition strength.
For \isotope[160-166]{Gd}, no experimental data are currently available; therefore, the present
results should be regarded as predictions of the CBS rotor model.

\subsection{Dy (Z=66)}
\label{sec:Dy}

\begin{figure*}[ht]
    \centering
    \includegraphics[width=0.75\textwidth]{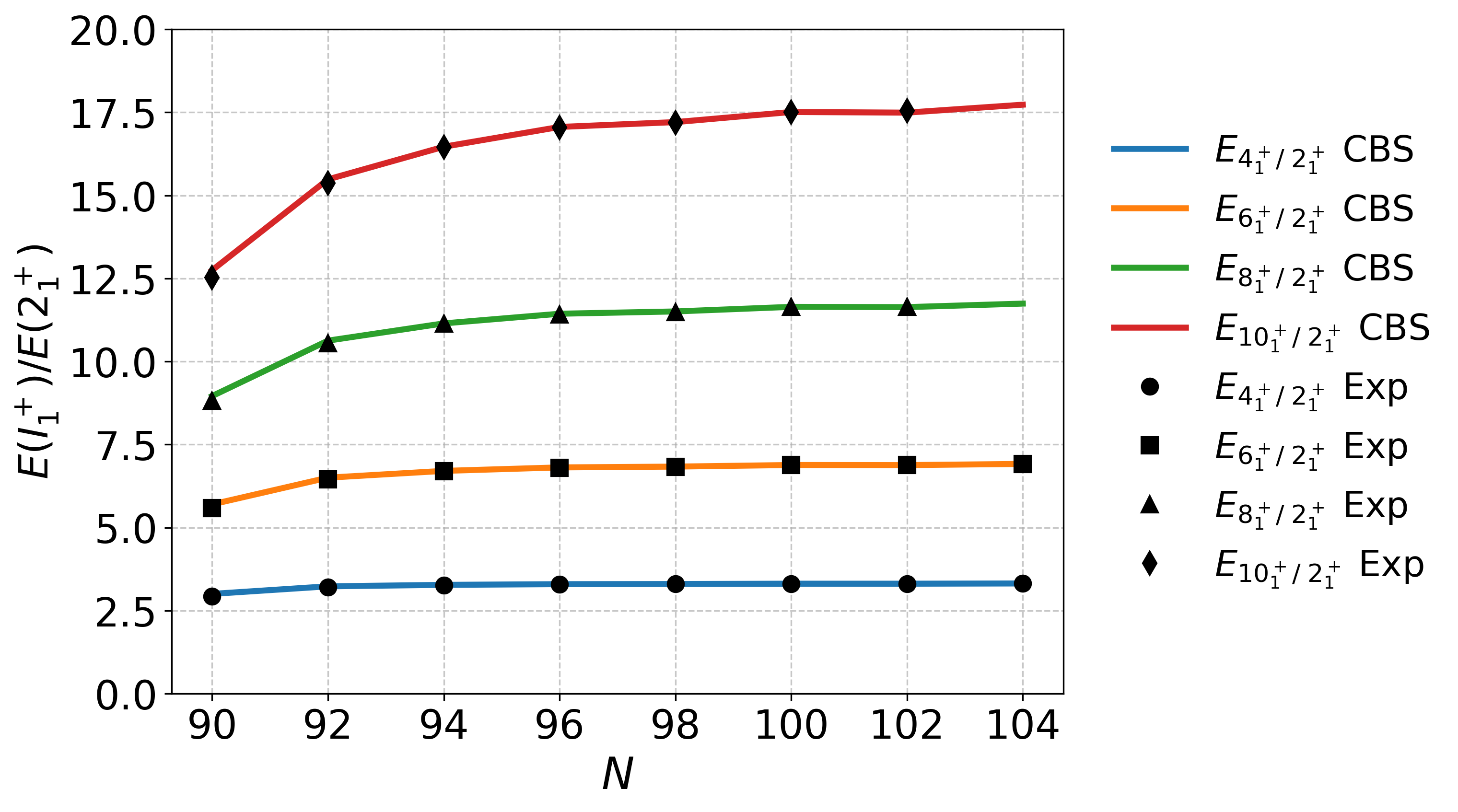}
    \caption{
Systematics of energy ratios
$E(I_1^+)/E(2_1^+)$
for the ground-state band of even--even Dy (Z=66) isotopes as a function of neutron number $N$. Solid lines indicate CBS predictions, while experimental values are shown as symbols. Experimental uncertainties are smaller than the symbol sizes.
}
    \label{fig:dy_energy_ratios}
\end{figure*}

\begin{figure*}[ht]
    \centering
    \includegraphics[width=0.7\textwidth]{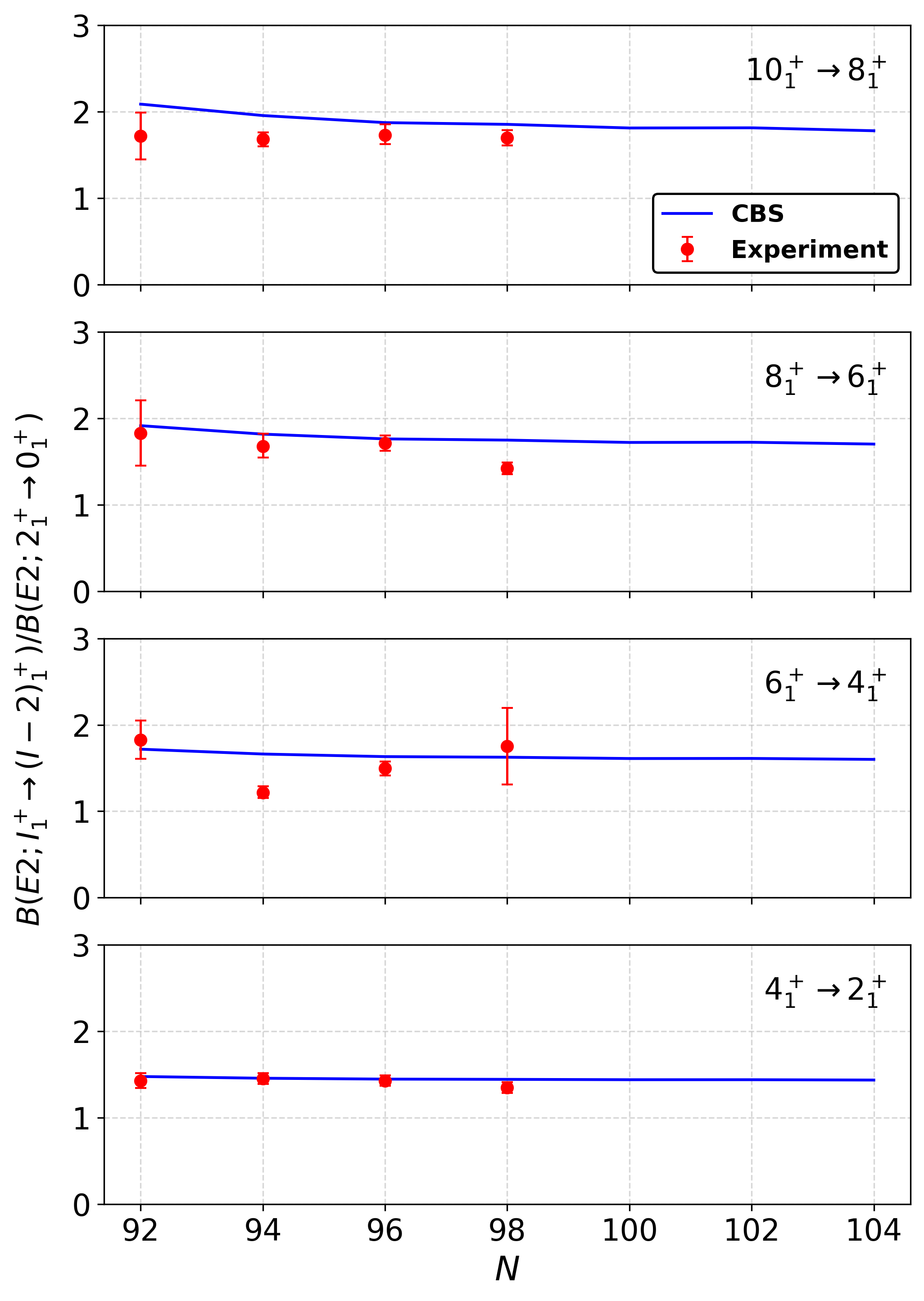}
    \caption{Systematics of reduced electric quadrupole transition ratios
$B(E2; I_1^+ \rightarrow I_1^+-2) / B(E2; 2_1^+ \rightarrow 0_1^+)$
for the ground-state band of even--even Dy (Z=66) isotopes as a function of neutron number $N$. Solid lines indicate CBS predictions, while experimental values are shown as symbols. Experimental data were retrieved from~\cite{REICH20122537, NICA20171, Nica:2021cny, Nica:2024ttv, PhysRevC.101.024313, singh_chen_2018}. When multiple measurements were available, the most recent values were used.}
    \label{fig:dy_BE2}
\end{figure*}

For the Dy isotopes, the CBS calculations shown in Fig.~\ref{fig:dy_energy_ratios} reproduce
the experimentally observed increase in rigidity with neutron number, reaching a maximum at
$N=104$. The saturation of the energy ratios with increasing neutron number is also evident,
indicating stabilization of the nuclear deformation, as further addition of neutrons after
$N=94$ does not significantly modify the collective structure. However, the same level of
agreement between experiment and CBS predictions is not achieved regarding the B(E2)s, plotted
in Fig.~\ref{fig:dy_BE2}. Since transition rates are quite sensitive to the detailed structure
and overlap of the wavefunctions, as well as to the specific form of the quadrupole operator,
$B(E2)$ observables probe finer structural features that are not entirely accounted for within
the simplified CBS framework.

\subsection{Er (Z=68)}
\label{sec:Er}

\begin{figure*}[ht]
    \centering
    \includegraphics[width=0.75\textwidth]{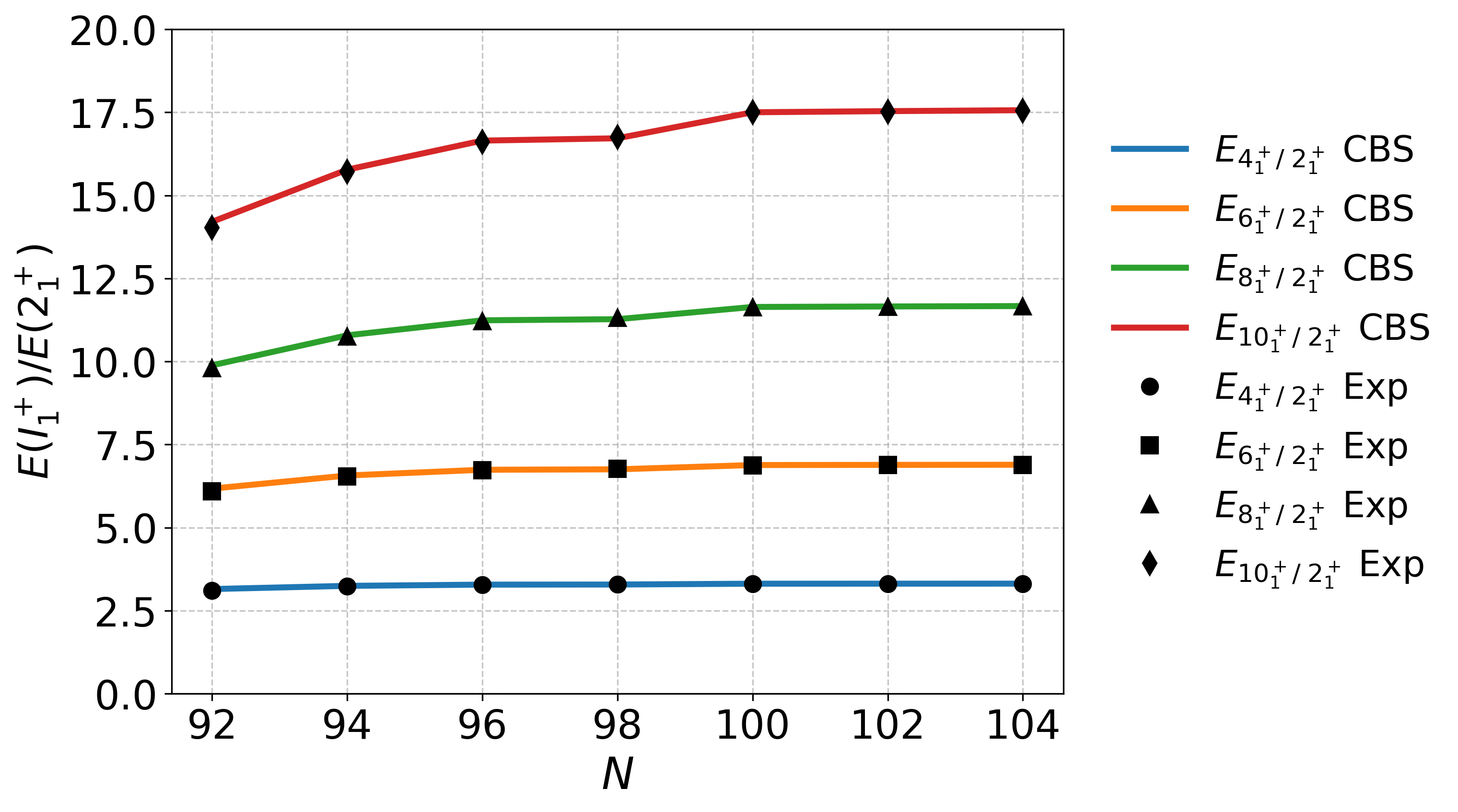}
    \caption{
Systematics of energy ratios
$E(I_1^+)/E(2_1^+)$
for the ground-state band of even--even Er (Z=68) isotopes as a function of neutron number $N$. Solid lines indicate CBS predictions, while experimental values are shown as symbols. Experimental uncertainties are smaller than the symbol sizes.
}
    \label{fig:er_energy_ratios}
\end{figure*}

\begin{figure*}[ht]
    \centering
    \includegraphics[width=0.7\textwidth]{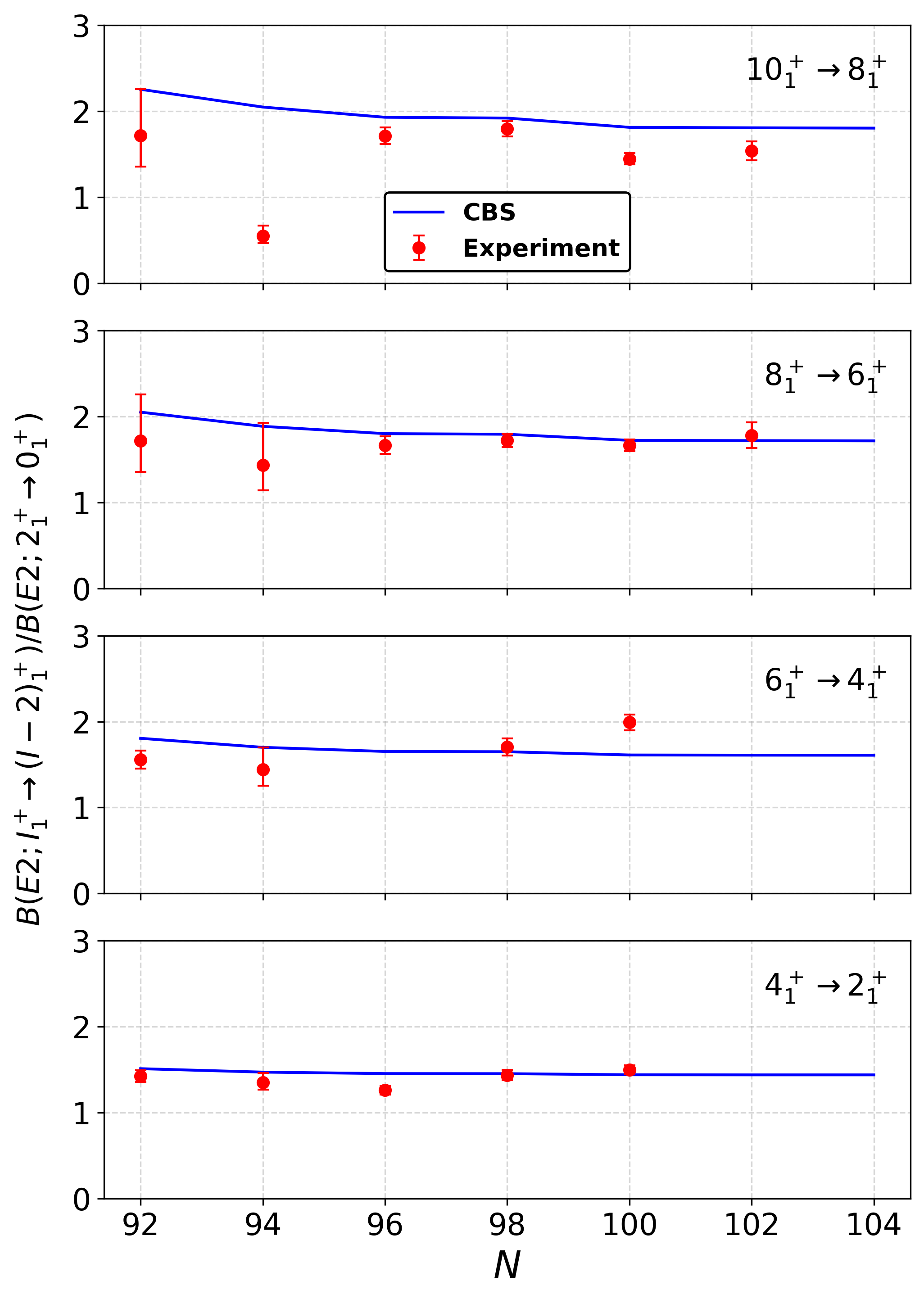}
    \caption{Systematics of reduced electric quadrupole transition ratios
$B(E2; I_1^+ \rightarrow I_1^+-2) / B(E2; 2_1^+ \rightarrow 0_1^+)$
for the ground-state band of even--even Er (Z=68) isotopes as a function of neutron number $N$. Solid lines indicate CBS predictions, while experimental values are shown as symbols. Experimental data were retrieved from~\cite{Nica:2021cny, Kocheva2026Er, singh_chen_2018, Baglin2008, Baglin2010, BAGLIN20181}. When multiple measurements were available, the most recent values were used.}
    \label{fig:er_BE2}
\end{figure*}

In Fig.~\ref{fig:er_energy_ratios}, saturation of the energy ratios of the ground-state
band is once again observed, particularly in the region $N=100-104$. Measurements of more
neutron-rich Er isotopes would be of considerable interest in order to determine whether
this saturation persists or whether the energy ratios decrease, indicating a deviation from
rigid-rotor behavior and a possible re-emergence of single-particle effects influencing the
ground-state deformation. In Fig.~\ref{fig:er_BE2}, many CBS predictions lie within the
experimental uncertainties, or at least close to them, indicating the collective character
of these transitions. However, as in most cases, the more neutron-rich Er isotopes, especially
\isotope[172]{Er} ($N=104$), lack experimental $B(E2)$ data. The predicted trend may therefore
serve as guidance for future experimental investigations. Finally, it should be noted that
the $B_{10/2}$ ratio for \isotope[162]{Er} ($N=94$), recently measured by Kocheva
\textit{et al.}~\cite{Kocheva2026Er}, is significantly lower than the general trend and
suggests a possible onset of backbending. Such effects lie beyond the scope of the CBS rotor
model, which does not account for quasiparticle alignments~\cite{PhysRevC.72.011303}.

\subsection{Yb (Z=70)}
\label{sec:Yb}

\begin{figure*}[ht]
    \centering
    \includegraphics[width=0.75\textwidth]{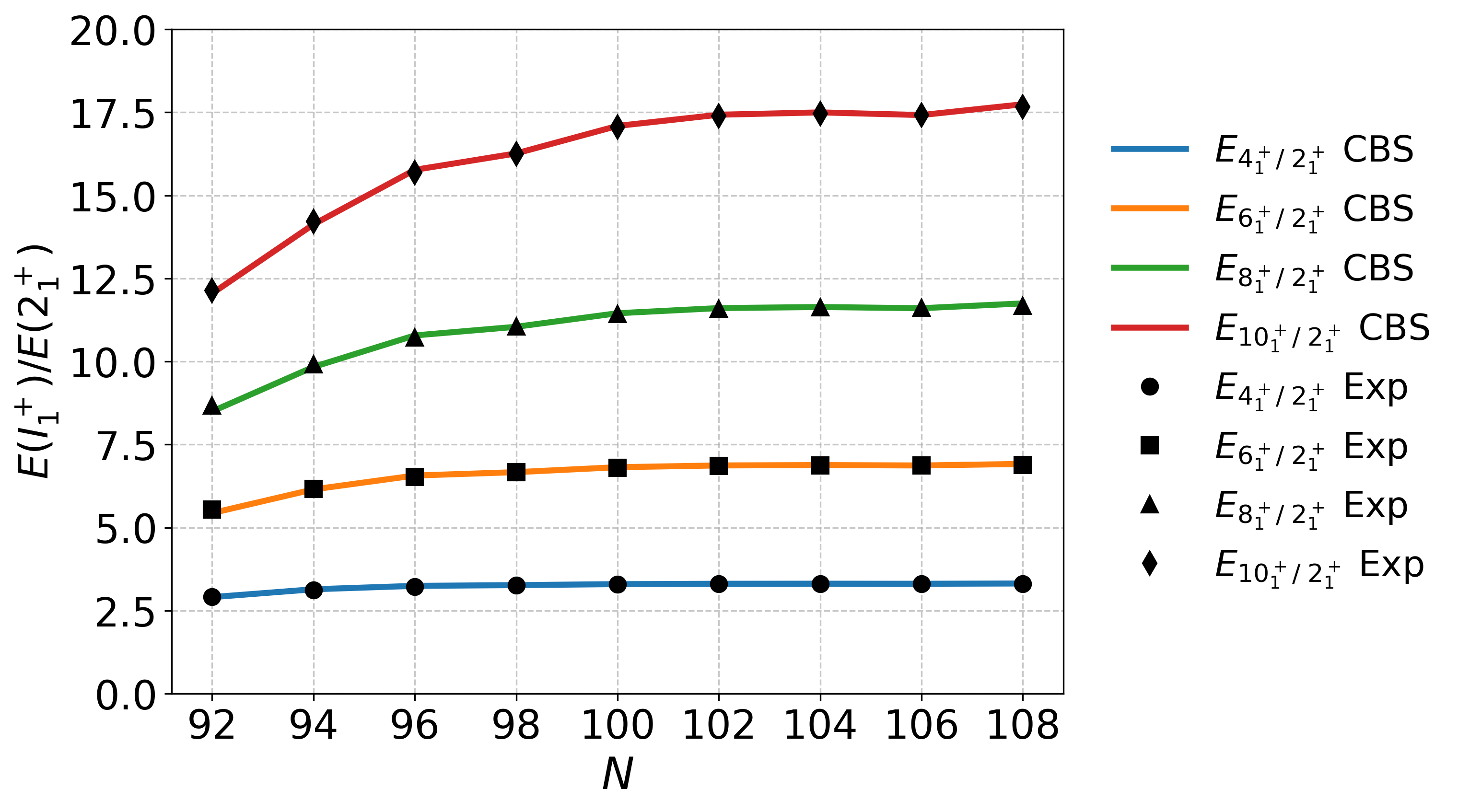}
    \caption{
Systematics of energy ratios
$E(I_1^+)/E(2_1^+)$
for the ground-state band of even--even Yb (Z=70) isotopes as a function of neutron number $N$. Solid lines indicate CBS predictions, while experimental values are shown as symbols. Experimental uncertainties are smaller than the symbol sizes.
}
    \label{fig:Yb_energy}
\end{figure*}

\begin{figure*}[ht]
    \centering
    \includegraphics[width=0.7\textwidth]{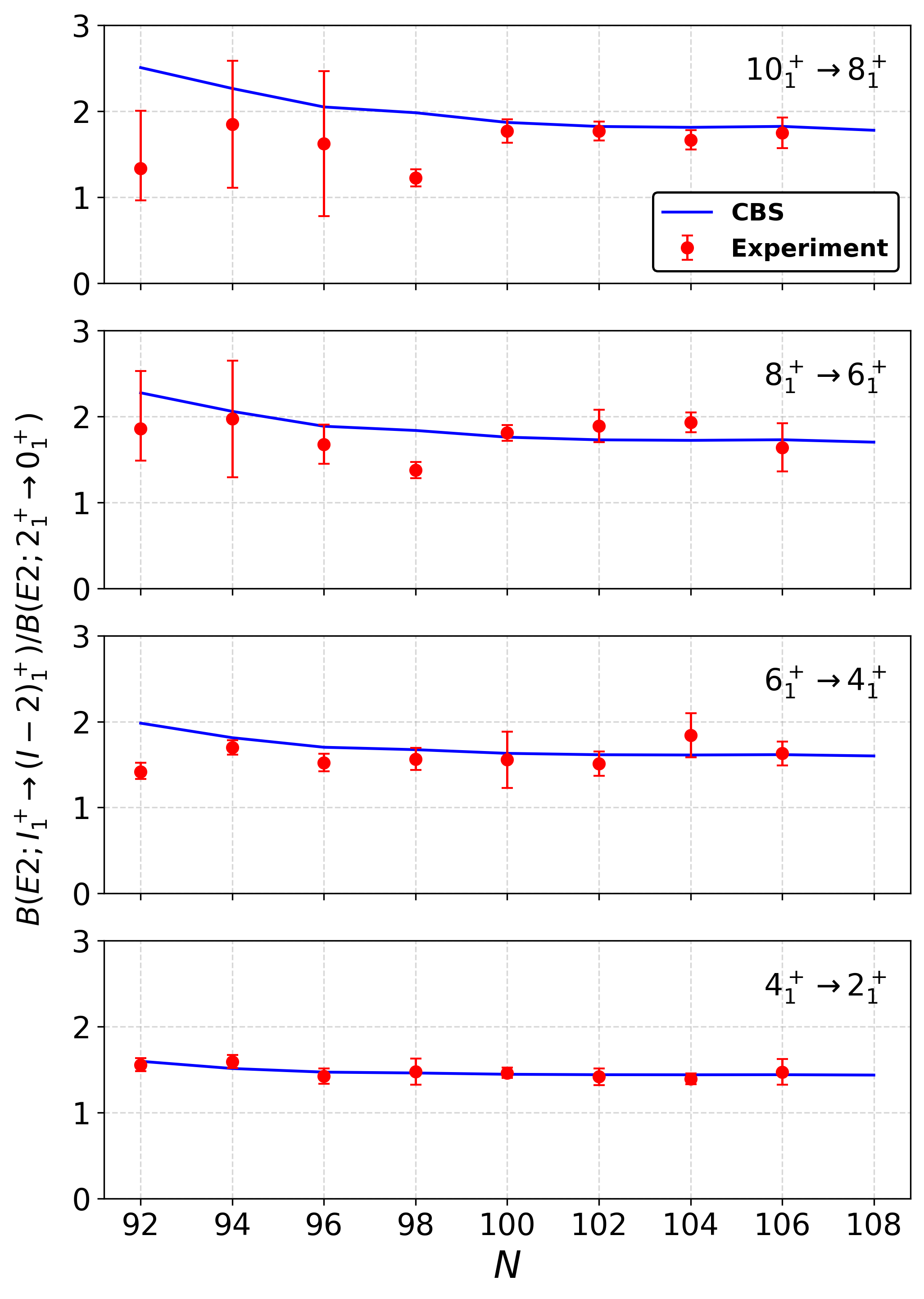}
    \caption{Systematics of reduced electric quadrupole transition ratios
$B(E2; I_1^+ \rightarrow I_1^+-2) / B(E2; 2_1^+ \rightarrow 0_1^+)$
for the ground-state band of even--even Yb (Z=70) isotopes as a function of neutron number $N$. Solid lines indicate CBS predictions, while experimental values are shown as symbols. Experimental data were retrieved from~\cite{Nica:2024ttv, singh_chen_2018, Baglin2008, PETKOV2017240, PhysRevC.95.034316, BAGLIN20181}. When multiple measurements were available, the most recent values were used.}
    \label{fig:Yb_BE2}
\end{figure*}

For the Yb isotopic chain, good agreement between the CBS predictions and the experimental
ground-state band energies is once again observed in Fig.~\ref{fig:Yb_energy}. The energy
ratios exhibit a steady increase with neutron number \(N\) and reach a maximum at
\isotope[178]{Yb} ($N=108$, $r_\beta$=0.509), reflecting the strengthening of rotational
rigidity with increasing neutron number~\cite{Koseoglou_2026, koseoglou_inprep}. It should
be noted that the CBS calculation based on the experimental data available in the
literature~\cite{Achterberg2009} yields a value of $r_\beta = 0.496$. However, as shown
in Fig.~\ref{fig:rb}, the use of the new measurements reported in \cite{koseoglou_inprep}
leads to $r_\beta = 0.509$, thereby establishing \isotope[178]{Yb} as the most rigid nucleus
in the rare-earth region. In addition, a substantial amount of experimental \(B(E2)\) data
is available, allowing for systematic comparison. As shown in Fig.~\ref{fig:Yb_BE2}, most
CBS predictions lie within the experimental uncertainties for the corresponding transition
ratios. However, noticeable deviations are observed for \isotope[168]{Yb} ($N=98$) in the
\(B_{8/2}\) and \(B_{10/2}\) ratios, where the CBS model fails to reproduce the observed
decrease in the B(E2) transition strengths. This drop suggests that the ground-state band
is no longer purely collective at higher spin, and that single-particle degrees of freedom
begin to play a more significant role. For \isotope[178]{Yb}, no experimental \(B(E2)\) data
are currently available, as lifetime measurements of the low-lying excited states in the
ground-state band remain experimentally challenging~\cite{Zyriliou:2022gkx}.

\subsection{Hf (Z=72)}
\label{sec:Hf}

\begin{figure*}[ht]
    \centering
    \includegraphics[width=0.75\textwidth]{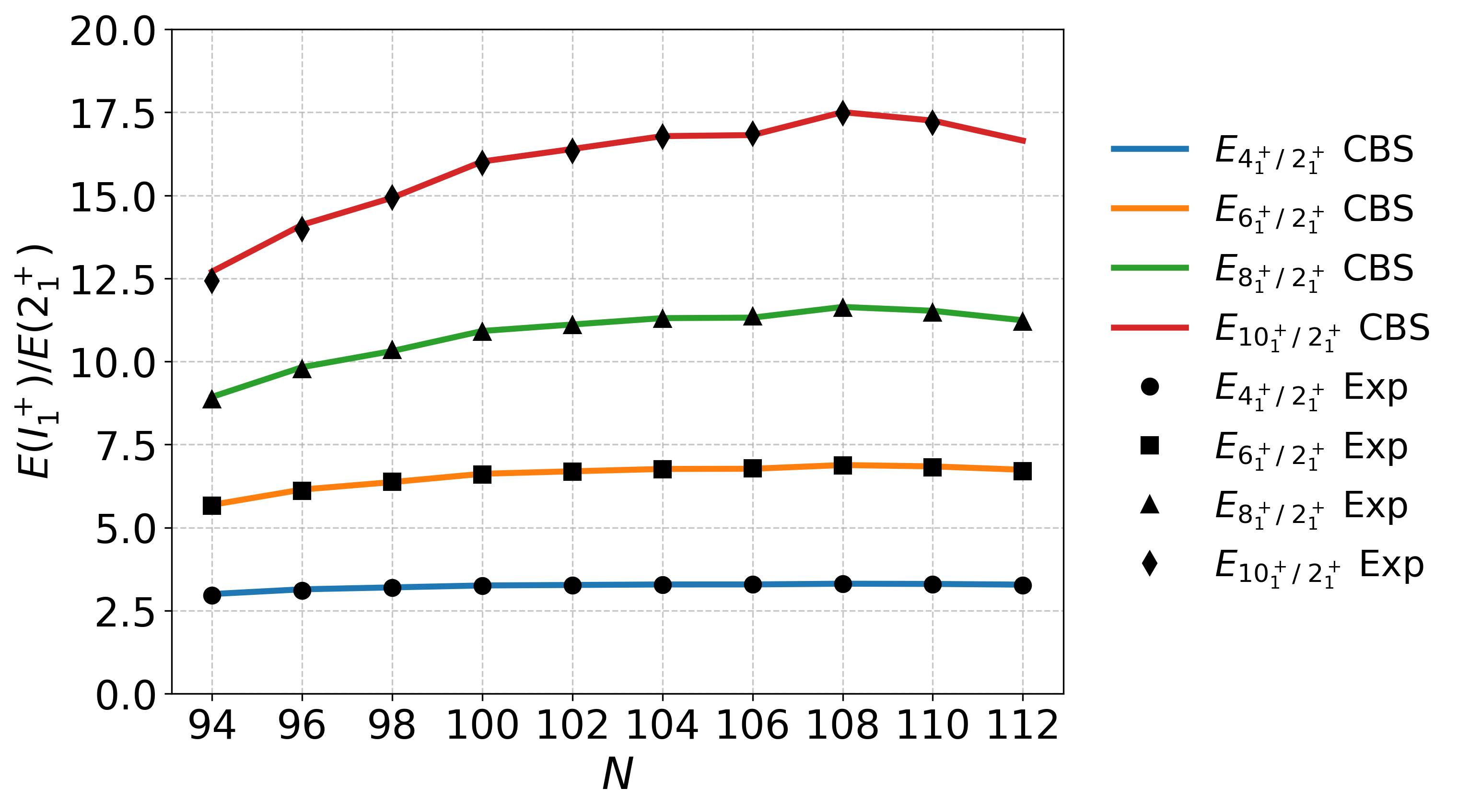}
    \caption{
Systematics of energy ratios
$E(I_1^+)/E(2_1^+)$
for the ground-state band of even--even Hf (Z=72) isotopes as a function of neutron number $N$. Solid lines indicate CBS predictions, while experimental values are shown as symbols. Experimental uncertainties are smaller than the symbol sizes.
}
    \label{fig:hf_energy}
\end{figure*}

\begin{figure*}[ht]
    \centering
    \includegraphics[width=0.7\textwidth]{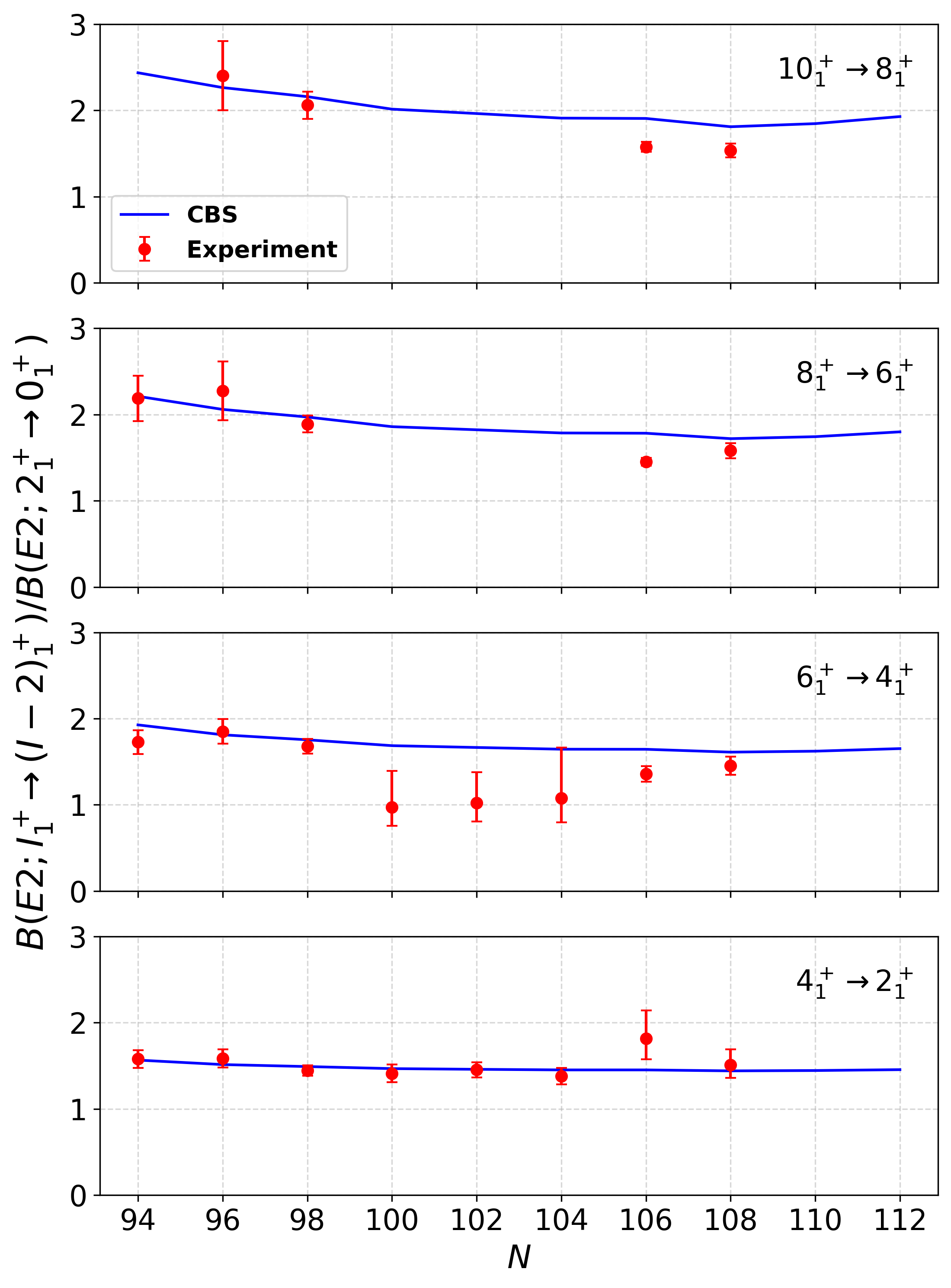}
    \caption{Systematics of reduced electric quadrupole transition ratios
$B(E2; I_1^+ \rightarrow I_1^+-2) / B(E2; 2_1^+ \rightarrow 0_1^+)$
for the ground-state band of even--even Hf (Z=72) isotopes as a function of neutron number $N$. Solid lines indicate CBS predictions, while experimental values are shown as symbols. Experimental data were retrieved from~\cite{Baglin2008, Baglin2010, BAGLIN20181, PhysRevC.91.044301, PhysRevC.99.024316}. When multiple measurements were available, the most recent values were used.}
    \label{fig:hf_BE2}
\end{figure*}

In Fig.~\ref{fig:hf_energy}, a characteristic peak in the energy ratios is observed at
$N=108$ (\isotope[108]{Hf}). For $N=108$, the $r_\beta$ parameter also reaches its highest
value ($0.463$) within the Hf isotopic chain, indicating close proximity to the rigid-rotor
limit. This maximum in rigidity at $N=108$ is also observed in the Yb, Hf, W, and Os
isotopic chains, suggesting a systematic structural effect. As recently
discussed~\cite{GUPTA2025123032}, $N=108$ appears to act as a transition point where nuclei
approach the SU(3) symmetry and subsequently evolve back toward the X(5) limit. This
structural evolution is consistent with the observed decrease in the energy ratios beyond
$N=108$. For the $B(E2)$ transitions, the CBS rotor model, following a purely collective
approach, predicts a minimum of the $B(E2)$ ratios at $N=108$, in analogy with the peak
observed in the energy ratios (see Fig.~\ref{fig:hf_BE2}). However, the available experimental
data do not fully support this behavior, with the exception of the low-lying $B_{4/2}$ ratio.
This deviation is not fully surprising, since quadrupole transition rates are influenced by
additional structural effects, and single-particle degrees of freedom may play a non-negligible
role, especially for transitions between higher energy levels.

\subsection{W (Z=74)}
\label{sec:W}

\begin{figure*}[ht]
    \centering
    \includegraphics[width=0.7\textwidth]{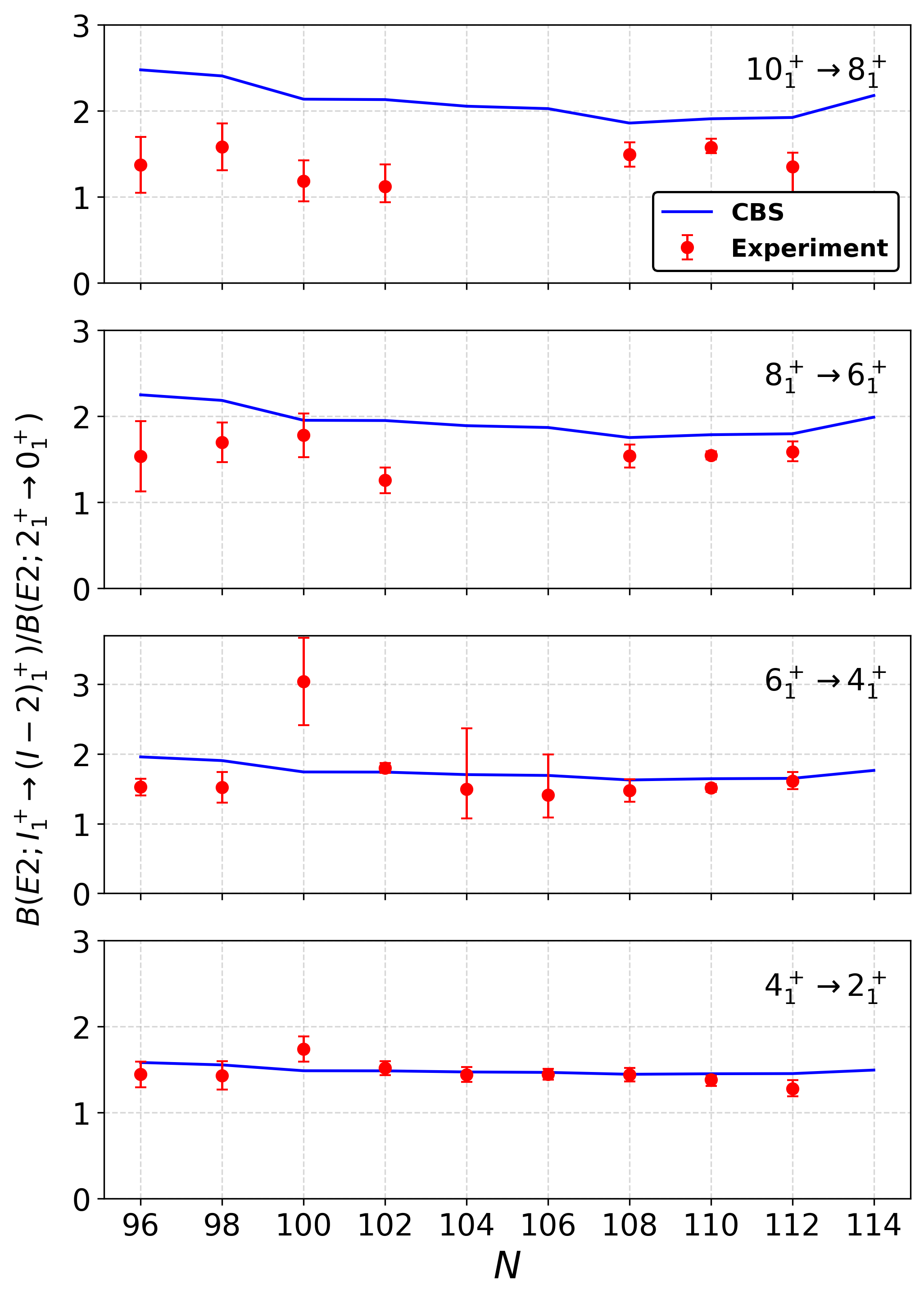}
    \caption{Systematics of reduced electric quadrupole transition ratios
$B(E2; I_1^+ \rightarrow I_1^+-2) / B(E2; 2_1^+ \rightarrow 0_1^+)$
for the ground-state band of even--even W (Z=74) isotopes as a function of neutron number $N$. Solid lines indicate CBS predictions, while experimental values are shown as symbols. Experimental data were retrieved from~\cite{BAGLIN20181, Singh1995, Browne1999, PhysRevC.106.024326, SINGH201521, BAGLIN2010275, BATCHELDER20221}. When multiple measurements were available, the most recent values were used.}
    \label{fig:w_BE2}
\end{figure*}

Similarly to the behavior observed in the Hf isotopes, the energy ratios in the W isotopic
chain exhibit a maximum at $N=108$, indicating the end of the saturation, followed by a
characteristic decrease (Fig.~\ref{fig:w_energy}). In Fig.~\ref{fig:w_BE2}, the CBS calculations
reproduce the overall experimental trends across all transitions, particularly for the
$4_1^+ \rightarrow 2_1^+$ and $6_1^+ \rightarrow 4_1^+$ ratios. For higher-spin transitions
($8_1^+ \rightarrow 6_1^+$ and $10_1^+ \rightarrow 8_1^+$), the model continues to capture
the general behavior, although some deviations are observed, especially at lower neutron
numbers where experimental uncertainties are larger. The CBS predictions exhibit a smooth
dependence on neutron number, indicating a gradual evolution of the rotational rigidity.
A systematic investigation of W isotopes can be found in~\cite{GUPTA2025123032}.

\subsection{Os (Z=76)}
\label{sec:Os}

\begin{figure*}[ht]
    \centering
    \includegraphics[width=0.75\textwidth]{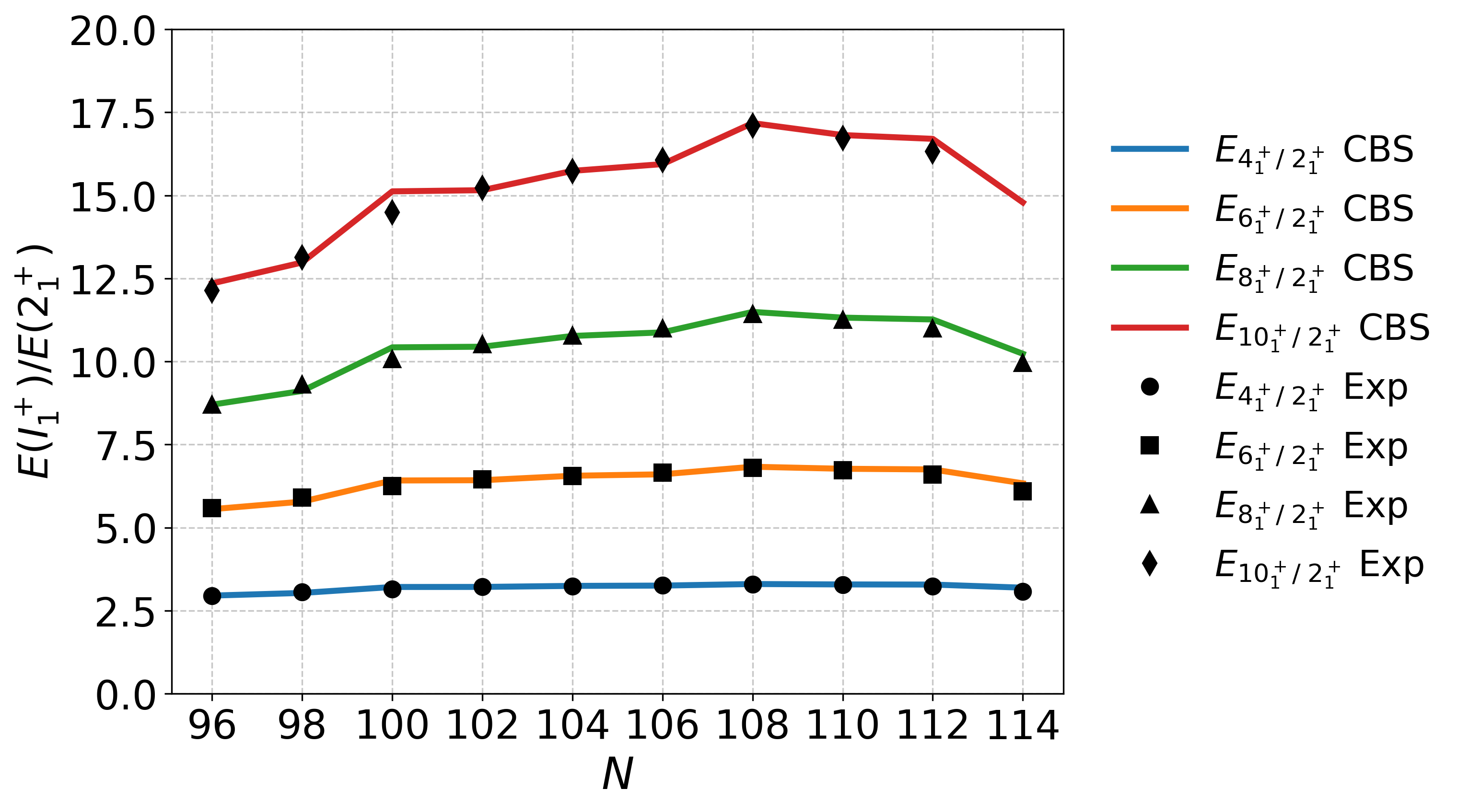}
    \caption{
Systematics of energy ratios
$E(I_1^+)/E(2_1^+)$
for the ground-state band of even--even W (Z=74) isotopes as a function of neutron number $N$. Solid lines indicate CBS predictions, while experimental values are shown as symbols. Experimental uncertainties are smaller than the symbol sizes.
}
    \label{fig:w_energy}
\end{figure*}

\begin{figure*}[ht]
    \centering
    \includegraphics[width=0.75\textwidth]{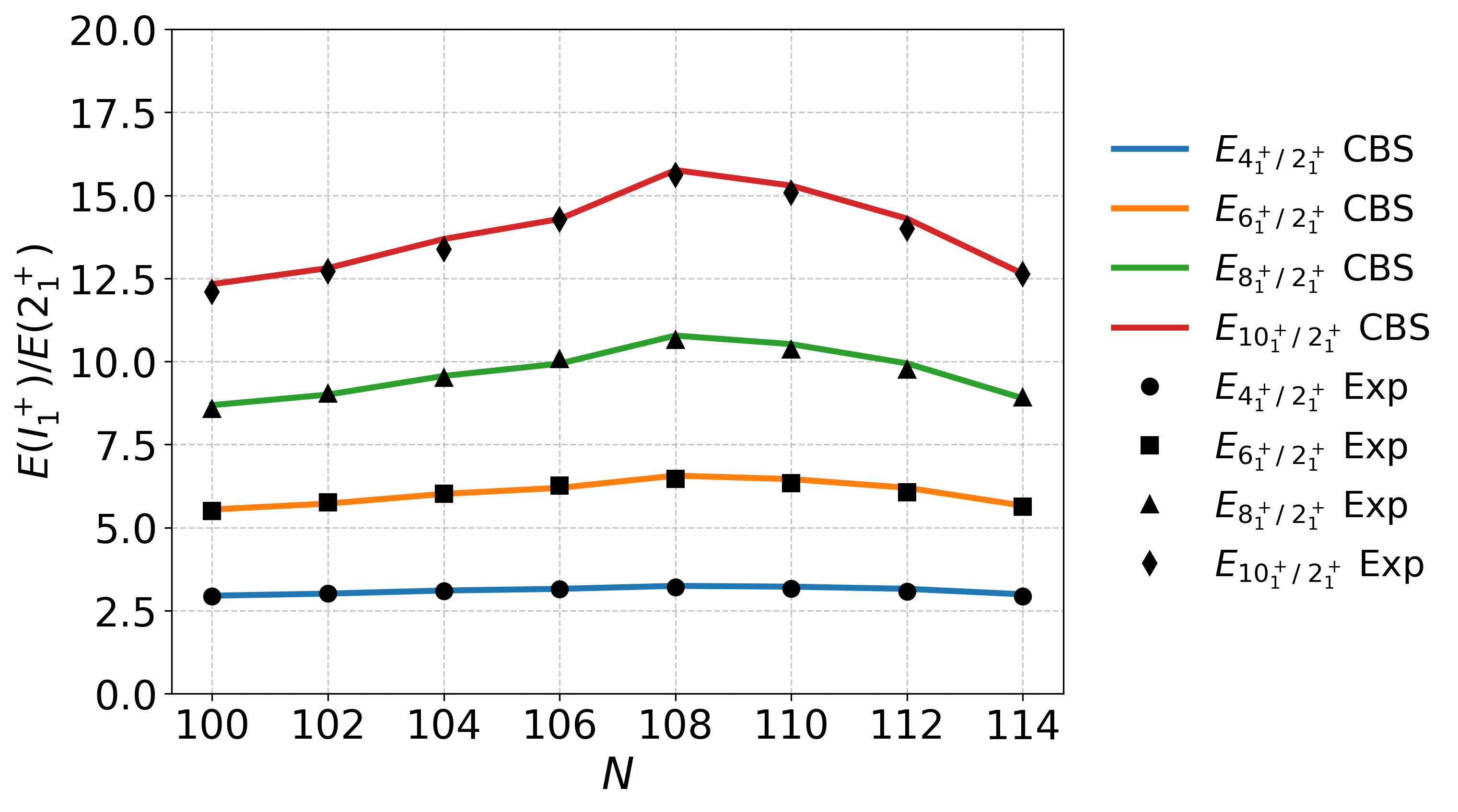}
    \caption{
Systematics of energy ratios
$E(I_1^+)/E(2_1^+)$
for the ground-state band of even--even Os (Z=76) isotopes as a function of neutron number $N$. Solid lines indicate CBS predictions, while experimental values are shown as symbols. Experimental uncertainties are smaller than the symbol sizes.
}
    \label{fig:os_energy}
\end{figure*}

\begin{figure*}[ht]
    \centering
    \includegraphics[width=0.7\textwidth]{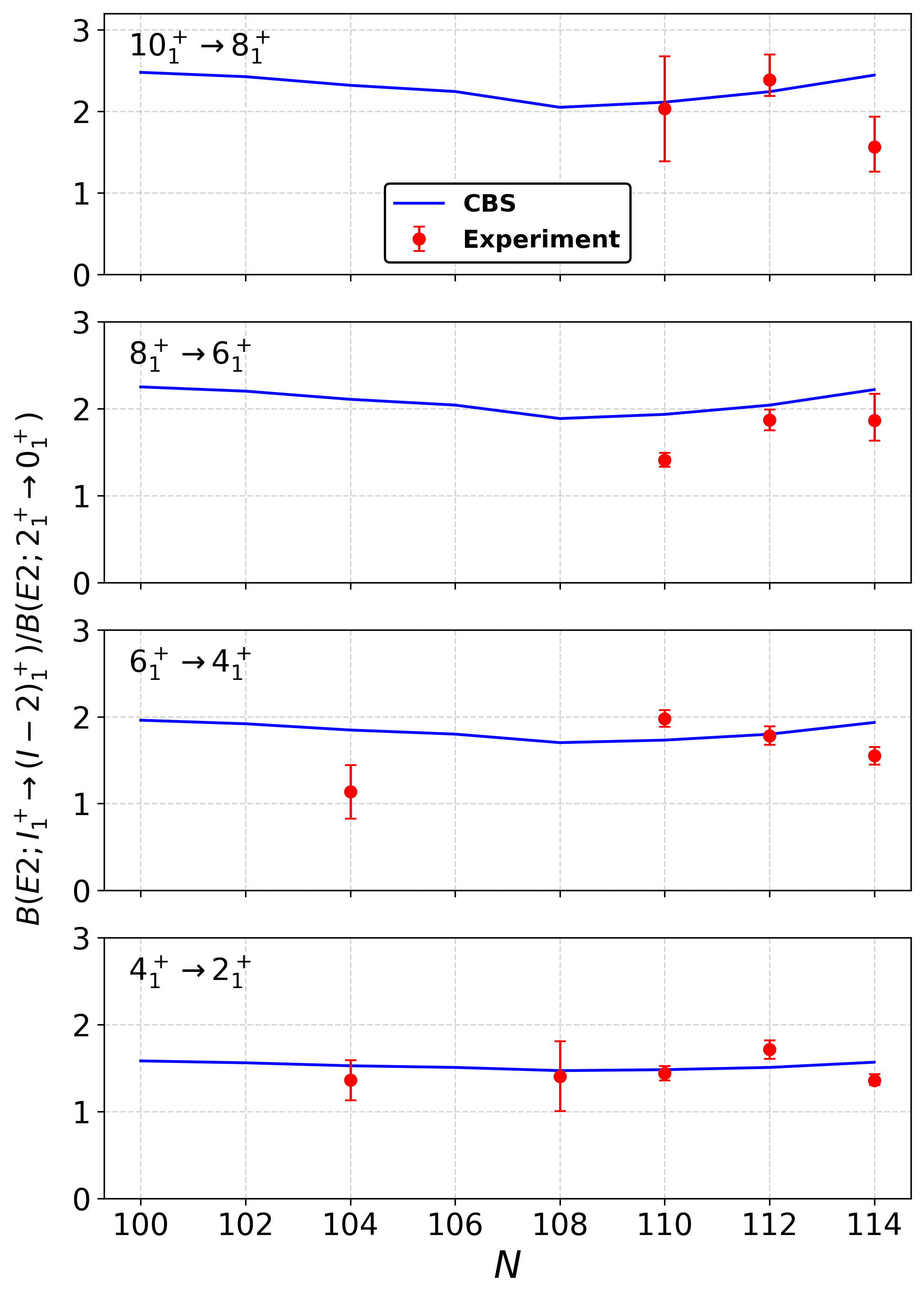}
    \caption{Systematics of reduced electric quadrupole transition ratios
$B(E2; I_1^+ \rightarrow I_1^+-2) / B(E2; 2_1^+ \rightarrow 0_1^+)$
for the ground-state band of even--even Os (Z=76) isotopes as a function of neutron number $N$. Solid lines indicate CBS predictions, while experimental values are shown as symbols. Experimental data were retrieved from~\cite{McCutchan2015, PhysRevC.108.024305, BAGLIN2010275, BATCHELDER20221, KONDEV20181, Singh:2020zfc}. When multiple measurements were available, the most recent values were used.}
    \label{fig:os_BE2}
\end{figure*}

The systematically smaller $r_\beta$ parameters and lower $R_{4/2}$ ratios observed in the
W and Os isotopic chains (see Figs.~\ref{fig:rb}, \ref{fig:w_energy}, and \ref{fig:os_energy})
can be attributed to their closer proximity to the $Z=82$ proton shell closure. As the proton
number approaches the shell closure at $Z=82$, the available valence space decreases, weakening
the quadrupole interactions responsible for deformation. This suppresses the overall rigidity,
which is reflected in reduced $R_{4/2}$ ratios. Within the CBS framework, this structural
effect is reflected in the smaller fitted values of the $r_\beta$ parameter, indicating
reduced rigidity compared to e.g. Er and Yb. Additionally, Fig.~\ref{fig:os_BE2} provides
guidance for future experimental studies, as only limited experimental data are currently
available for \isotope[176-184]{Os}.

\subsection{Identifying possible X(5) candidates}
\label{sec:X5}

This section aims to identify isotopes that may be described within the $X(5)$ critical-point
symmetry, namely nuclei located in the transitional region between spherical shapes and rigid
rotors. Within the CBS interpretation, small values of the parameter $r_\beta$ correspond to
nuclei lying closer to the $X(5)$ limit (see Fig.~\ref{fig:rb} and the related discussion).
Among the nuclei traditionally discussed as possible X(5) candidates in the rare-earth region
are \isotope[150]{Nd}, \isotope[152]{Sm}, \isotope[154]{Gd}, and
\isotope[156]{Dy}~\cite{Caprio2005StructureOC,PhysRevC.96.034321}. These nuclei belong to the
well-known $N=90$ isotones, which are commonly associated with the spherical-to-axially-deformed
phase-transitional region. In the present work, the corresponding fitted CBS values of $r_\beta$
are found to be 0.098, 0.199, 0.203, and 0.139, respectively, further supporting their proximity
to the X(5) limit. Moreover, several additional nuclei have been examined in the literature as
possible X(5) candidates, such as \isotope[128]{Ce}, \isotope[162]{Yb}, \isotope[166]{Hf}, and \isotope[176]{Os}, with corresponding $r_\beta$ values of 0.138, 0.036, 0.136, and 0.098, respectively~\cite{balabanski,PhysRevC.69.034334,mccucthan,RevModPhys.82.2155}. 

\section{Conclusions}
\label{sec:concl}

In the present work, a systematic study involving the CBS model was undertaken in
the even-even isotopes from Ce ($Z=58$) to Os ($Z=76$). Ratios of level energies
and reduced matrix elements, which generally serve as useful guides to understand
structure evolution across the nuclear chart, have been calculated.

The results showcase the efficiency of the CBS model to describe the collective
behavior of these nuclei in the ground-state and beta bands. They also highlight
the existence of single-particle components in the wavefunction of some of these
isotopes, stressing the need to expand the existing measurements to the outskirts
of nuclear stability.

It is a core belief of the authors that the complete set of calculations presented
in the tables and figures of this work may serve well researchers interested in
exploring the nuclear structure properties in the rare-earth region. The results
can be compared with predictions of other established models performing well in the
same mass region (e.g. algebraic models) or be used to support future experimental
endeavors. To that end, Table~\ref{tab:be2_ratios} highlights several missing B(E2)
ratios, serving as a map to explore uncharted territories.

\section*{Acknowledgments}

\noindent PK acknowledges support by the DFG, Grant No 539757749.

\bibliographystyle{elsarticle-num-names} 
\bibliography{biblio_rev1}

\clearpage

\section*{Figures}


\clearpage

\TableExplanation

\section*{Table~A. Comparison between theoretical (CBS) and experimental energy ratios.}

Comparison between experimental (\emph{Exp.}) data and theoretical predictions obtained within the Confined $\beta$-Soft (CBS) rotor model for even--even nuclei in the rare-earth region.
The table presents ratios of excitation energies in the ground-state band,
\[
E(I_1^+)/E(2_1^+),
\]
where $I_1^+$ denotes the spin of the excited state. 
For each observable, the CBS result is listed alongside the corresponding experimental value, where available.
The parameter $r_\beta$ denotes the confinement parameter of the CBS model and is determined independently for each nucleus.

\begin{center}
\begin{tabular}{ll}
Isotope & Even--even nucleus under consideration \\[4pt]
$r_\beta$ & CBS confinement parameter determined from the fit \\[4pt]
$E(4_1^+)/E(2_1^+)$, $E(6_1^+)/E(2_1^+)$, $\dots$ & Energy ratios for the ground-state band \\[4pt]
CBS & Predictions obtained from the CBS model \\[4pt]
Exp. & Experimental values taken from evaluated data (ENSDF)
\end{tabular}
\end{center}

\clearpage

\section*{Table~B. Comparison between theoretical (CBS) and experimental $B(E2)$ transition ratios.}

Comparison between experimental (\emph{Exp.}) data and theoretical predictions obtained within the Confined $\beta$-Soft (CBS) rotor model for even--even nuclei in the rare-earth region.
The table presents reduced electric quadrupole transition-probability ratios in the ground-state band,
\[
\frac{B(E2; I_1^+ \rightarrow (I_1-2)_1^+)}{B(E2; 2_1^+ \rightarrow 0_1^+)},
\]
where $I_1^+$ denotes the spin of the initial state ($I\neq 2$). 
For each observable, the CBS result is listed alongside the corresponding experimental value, where available.

\begin{center}
\begin{tabular}{ll}
Isotope & Even--even nucleus under consideration \\[4pt]
$r_\beta$ & CBS confinement parameter determined from the fit \\[4pt]
$B(E2; I_1^+ \rightarrow (I_1-2)_1^+)/B(E2; 2_1^+ \rightarrow 0_1^+)$
& Reduced transition-probability ratios for the ground-state band \\[4pt]
CBS & Predictions obtained from the CBS model \\[4pt]
Exp. & Experimental values taken from evaluated data (ENSDF)
\end{tabular}
\end{center}

\section*{Table~C. CBS predictions for \(\beta\)-band excitation energies and $B(E2)$ transition ratios.}

Predictions obtained within the Confined \(\beta\)-Soft (CBS) rotor model for even--even nuclei in the rare-earth region.
The table presents excitation energies (in keV) of the \(\beta\)-band,
\[
E(I_\beta^+), \qquad I_\beta^+ = 0_\beta^+,\,2_\beta^+,\,4_\beta^+,\,6_\beta^+,
\]
where \(I_\beta^+\) denotes the spin/parity of states belonging to the \(\beta\)-band.
For each nucleus, the CBS prediction is listed.
The parameter \(r_\beta\) denotes the confinement parameter of the CBS model and is determined independently for each nucleus.
In addition, the table includes CBS predictions for intraband reduced electric quadrupole transition ratios within the \(\beta\) band,
\[
\frac{B(E2;4_\beta^+\!\rightarrow 2_\beta^+)}{B(E2;2_\beta^+\!\rightarrow 0_\beta^+)},
\qquad
\frac{B(E2;6_\beta^+\!\rightarrow 4_\beta^+)}{B(E2;2_\beta^+\!\rightarrow 0_\beta^+)}.
\]

\begin{center}
\begin{tabular}{ll}
Isotope & Even--even nucleus under consideration \\[4pt]
$r_\beta$ & CBS confinement parameter determined from the fit \\[4pt]
$E(I_\beta^+)$ & CBS excitation energies (keV) of \(\beta\)-band states \\[4pt]
CBS & Predictions obtained from the CBS model \\[4pt]
CBS \(B(E2)\) ratios & Intraband \(\beta\)-band transition ratios predicted by the CBS model
\end{tabular}
\end{center}


\newcommand{\IsoRow}[1]{%
#1 & 
 & 
 & & & & & & & 
 & & & & & & & 
\\ \hline
}

\newcommand{\IsoRowFilled}[2]{%
#1 & #2 \\ \hline
}

\clearpage

\scriptsize
\setlength{\LTpre}{0pt}
\setlength{\LTpost}{0pt}

\begin{longtable}{|c|c|c|c|c|c|c|c|c|c|}

\caption{Energy ratios for even--even nuclei in the ground-state band. CBS results are compared with experimental values.}
\\\hline

\label{tab:energy_ratios}
\multirow{3}{*}{Isotope}
& \multirow{3}{*}{$r_\beta$}
& \multicolumn{8}{c|}{Energy ratios} \\ \cline{3-10}

&
& \multicolumn{2}{|c|}{$\dfrac{E(4_1^+)}{E(2_1^+)}$}
& \multicolumn{2}{|c|}{$\dfrac{E(6_1^+)}{E(2_1^+)}$}
& \multicolumn{2}{|c|}{$\dfrac{E(8_1^+)}{E(2_1^+)}$}
& \multicolumn{2}{|c|}{$\dfrac{E(10_1^+)}{E(2_1^+)}$}
\\ \cline{3-10}

&
& CBS & Exp
& CBS & Exp
& CBS & Exp
& CBS & Exp \\
\hline
\endfirsthead

\multicolumn{10}{c}{\tablename\ \thetable\ -- continued}\\
\hline
\multirow{3}{*}{Isotope}
& \multirow{3}{*}{$r_\beta$}
& \multicolumn{8}{c|}{Energy ratios} \\ \cline{3-10}
&
& \multicolumn{2}{c|}{$\dfrac{E(4_1^+)}{E(2_1^+)}$}
& \multicolumn{2}{c|}{$\dfrac{E(6_1^+)}{E(2_1^+)}$}
& \multicolumn{2}{c|}{$\dfrac{E(8_1^+)}{E(2_1^+)}$}
& \multicolumn{2}{c|}{$\dfrac{E(10_1^+)}{E(2_1^+)}$}
\\ \cline{3-10}
&
& CBS & Exp
& CBS & Exp
& CBS & Exp
& CBS & Exp \\
\hline
\endhead

\hline
\multicolumn{10}{r}{continued on next page}\\
\endfoot

\hline
\endlastfoot


\IsoRowFilled{\ce{^{122}Ce}}{0.292 & 3.219 & {3.187} & 6.447 & {6.394} & 10.436 & {10.474} & 15.242 & {15.298}}
\IsoRowFilled{\ce{^{124}Ce}}{0.280 & 3.207 & {3.156} & 6.396 & {6.285} & 10.377 & {10.223} & 15.036 & {14.799}}
\IsoRowFilled{\ce{^{126}Ce}}{0.210 & 3.116 & {3.061} & 6.052 & 5.985 & 9.636 & 9.585 & 13.802 & 13.639}
\IsoRowFilled{\ce{^{128}Ce}}{0.138 & 3.003 & {2.930} & 5.694 & 5.587 & 8.952 & 8.785 & 12.732 & 12.219 }
\IsoRowFilled{\ce{^{150}Ce}}{0.272 & 3.198 & {3.152} & 6.363 & {6.252} & 10.300 & {10.130} & 14.903 & {14.666}}
\IsoRowFilled{\ce{^{152}Ce}}{0.364 & 3.274 & {3.251} & 6.697 & {6.632} & 11.119 & {11.010} & 16.409 & {16.269}}
\IsoRowFilled{\ce{^{154}Ce}}{0.410 & 3.295 & {3.307} & 6.800 & 6.823 & 11.404 & 11.440 & 16.989 & --- }

\hline \hline

\IsoRowFilled{\ce{^{128}Nd}}{0.273 & 3.199 & {3.176} & 6.367 & {6.344} & 10.308 & {10.303} & 14.918 & {14.863}}
\IsoRowFilled{\ce{^{130}Nd}}{0.188 & 3.083 & {3.052} & 5.940 & {5.901} & 9.415 & {9.350} & 13.450 & {13.206}}
\IsoRowFilled{\ce{^{150}Nd}}{0.098 & 2.950 & {2.926} & 5.547 & {5.530} & 8.687 & {8.673} & 12.331 & {12.276}}
\IsoRowFilled{\ce{^{152}Nd}}{0.382 & 3.283 & {3.267} & 6.743 & {6.686} & 11.243 & {11.135} & 16.657 & {16.522}}
\IsoRowFilled{\ce{^{154}Nd}}{0.424 & 3.300 & {3.296} & 6.826 & {6.800} & 11.477 & {11.421} & 17.143 & {17.077}}
\IsoRowFilled{\ce{^{156}Nd}}{0.454 & 3.309 & {3.306} & 6.872 & {6.853} & 11.611 & {11.579} & 17.434 & {17.396}}
\IsoRowFilled{\ce{^{158}Nd}}{0.437 & 3.304 & 3.302 & 6.849 & 6.844 & 11.543 & --- & 17.281 & --- }

\hline \hline

\IsoRowFilled{\ce{^{132}Sm}}{0.283 & 3.210 & {3.183} & 6.410 & {6.359} & 10.407 & {10.336} & 15.092 & {14.977}}
\IsoRowFilled{\ce{^{152}Sm}}{0.199 & 3.099 & {3.009} & 5.994 & {5.803} & 9.521 & {9.238} & 13.617 & {13.212}}
\IsoRowFilled{\ce{^{154}Sm}}{0.363 & 3.273 & {3.255} & 6.693 & {6.637} & 11.109 & {11.011} & 16.389 & {16.263}}
\IsoRowFilled{\ce{^{156}Sm}}{0.443 & 3.306 & {3.290} & 6.857 & {6.814} & 11.566 & {11.484} & 17.336 & {17.230}}
\IsoRowFilled{\ce{^{158}Sm}}{0.460 & 3.311 & {3.300} & 6.880 & {6.845} & 11.632 & {11.572} & 17.482 & {17.397}}
\IsoRowFilled{\ce{^{160}Sm}}{0.457 & 3.310 & {3.293} & 6.876 & {6.829} & 11.622 & {11.538} & 17.459 & {17.340}}
\IsoRowFilled{\ce{^{162}Sm}}{0.448 & 3.307 & 3.304 & 6.863 & --- & 11.585 & --- & 17.376 & ---}
\IsoRowFilled{\ce{^{164}Sm}}{0.394 & 3.289 & 3.259 & 6.767 & 6.771 & 11.318 & --- & 16.804 & ---}

\hline \hline

\IsoRowFilled{\ce{^{154}Gd}}{0.203 & 3.105 & {3.014} & 6.014 & {5.833} & 9.561 & {9.300} & 13.681 & {13.299}}
\IsoRowFilled{\ce{^{156}Gd}}{0.339 & 3.258 & {3.239} & 6.622 & {6.572} & 10.924 & {10.853} & 16.030 & {15.921}}
\IsoRowFilled{\ce{^{158}Gd}}{0.415 & 3.297 & {3.289} & 6.810 & {6.781} & 11.432 & {11.372} & 17.049 & {16.972}}
\IsoRowFilled{\ce{^{160}Gd}}{0.441 & 3.306 & {3.301} & 6.854 & {6.841} & 11.558 & {11.541} & 17.318 & {17.288}}
\IsoRowFilled{\ce{^{162}Gd}}{0.440 & 3.305 & {3.307} & 6.852 & {6.851} & 11.553 & {11.553} & 17.308 & {17.306}}
\IsoRowFilled{\ce{^{164}Gd}}{0.476 & 3.314 & {3.294} & 6.898 & {6.861} & 11.687 & {11.624} & 17.605 & {17.509}}
\IsoRowFilled{\ce{^{166}Gd}}{0.455 & 3.309 & {3.297} & 6.873 & {6.851} & 11.613 & {---} & 17.440 & {---}}

\hline \hline

\IsoRowFilled{\ce{^{156}Dy}}{0.139 & 3.004 & 2.934 & 5.699 & 5.592 & 8.960 & 8.821 & 12.748 & 12.522}
\IsoRowFilled{\ce{^{158}Dy}}{0.306 & 3.232 & {3.206} & 6.503 & {6.448} & 10.628 & {10.551} & 15.483 & {15.369}}
\IsoRowFilled{\ce{^{160}Dy}}{0.368 & 3.276 & {3.269} & 6.707 & {6.695} & 11.147 & {11.137} & 16.465 & {16.451}}
\IsoRowFilled{\ce{^{162}Dy}}{0.416 & 3.297 & {3.294} & 6.812 & {6.803} & 11.438 & {11.422} & 17.061 & {17.049}}
\IsoRowFilled{\ce{^{164}Dy}}{0.430 & 3.302 & {3.300} & 6.836 & {6.831} & 11.507 & {11.492} & 17.207 & {17.183}}
\IsoRowFilled{\ce{^{166}Dy}}{0.464 & 3.311 & {3.311} & 6.884 & {6.883} & 11.645 & {11.646} & 17.511 & {17.511}}
\IsoRowFilled{\ce{^{168}Dy}}{0.461 & 3.311 & {3.313} & 6.881 & {6.884} & 11.637 & {11.651} & 17.492 & {17.546}}
\IsoRowFilled{\ce{^{170}Dy}}{0.495 & 3.318 & {3.321} & 6.917 & {6.920} & 11.743 & {---} & 17.731 & {---}}

\hline \hline

\IsoRowFilled{\ce{^{160}Er}}{0.233 & 3.150 & {3.103} & 6.173 & {6.098} & 9.886 & {9.796} & 14.207 & {14.032}}
\IsoRowFilled{\ce{^{162}Er}}{0.323 & 3.246 & {3.231} & 6.568 & {6.534} & 10.788 & {10.751} & 15.776 & {15.710}}
\IsoRowFilled{\ce{^{164}Er}}{0.382 & 3.283 & {3.277} & 6.741 & {6.722} & 11.240 & {11.209} & 16.650 & {16.613}}
\IsoRowFilled{\ce{^{166}Er}}{0.387 & 3.286 & {3.289} & 6.754 & {6.772} & 11.274 & {11.312} & 16.720 & {16.756}}
\IsoRowFilled{\ce{^{168}Er}}{0.463 & 3.311 & {3.308} & 6.883 & {6.875} & 11.642 & {11.633} & 17.503 & {17.499}}
\IsoRowFilled{\ce{^{170}Er}}{0.467 & 3.312 & {3.310} & 6.888 & {6.880} & 11.657 & {11.644} & 17.537 & {17.520}}
\IsoRowFilled{\ce{^{172}Er}}{0.471 & 3.313 & {3.314} & 6.892 & {6.886} & 11.669 & {11.662} & 17.565 & {17.556}}

\hline \hline

\IsoRowFilled{\ce{^{162}Yb}}{0.036 & 2.907 & {2.923} & 5.437 & {5.543} & 8.497 & {8.670} & 12.047 & {12.141}}
\IsoRowFilled{\ce{^{164}Yb}}{0.229 & 3.143 & {3.126} & 6.150 & {6.163} & 9.837 & {9.916} & 14.127 & {14.222}}
\IsoRowFilled{\ce{^{166}Yb}}{0.323 & 3.246 & {3.227} & 6.567 & {6.525} & 10.785 & {10.728} & 15.770 & {15.687}}
\IsoRowFilled{\ce{^{168}Yb}}{0.354 & 3.268 & {3.266} & 6.669 & {6.671} & 11.045 & {11.057} & 16.264 & {16.255}}
\IsoRowFilled{\ce{^{170}Yb}}{0.419 & 3.298 & {3.293} & 6.818 & {6.805} & 11.453 & {11.435} & 17.092 & {17.061}}
\IsoRowFilled{\ce{^{172}Yb}}{0.454 & 3.309 & {3.305} & 6.871 & {6.857} & 11.608 & {11.585} & 17.429 & {17.396}}
\IsoRowFilled{\ce{^{174}Yb}}{0.462 & 3.311 & {3.309} & 6.882 & {6.877} & 11.640 & {11.636} & 17.499 & {17.467}}
\IsoRowFilled{\ce{^{176}Yb}}{0.453 & 3.309 & {3.310} & 6.870 & {6.873} & 11.604 & {11.613} & 17.419 & {17.424}}
\IsoRowFilled{\ce{^{178}Yb}}{0.496 & 3.317 & {3.310} & 6.917 & {6.882} & 11.749 & {11.682} & 17.744 & {17.661}}

\hline \hline

\IsoRowFilled{\ce{^{166}Hf}}{0.136 & 3.000 & {2.966} & 5.685 & {5.655} & 8.936 & {8.864} & 12.707 & {12.431}}
\IsoRowFilled{\ce{^{168}Hf}}{0.228 & 3.142 & {3.109} & 6.146 & {6.100} & 9.830 & {9.778} & 14.116 & {13.989}}
\IsoRowFilled{\ce{^{170}Hf}}{0.274 & 3.200 & {3.194} & 6.370 & {6.380} & 10.316 & {10.353} & 14.931 & {14.931}}
\IsoRowFilled{\ce{^{172}Hf}}{0.339 & 3.258 & {3.248} & 6.620 & {6.598} & 10.919 & {10.895} & 16.021 & {15.974}}
\IsoRowFilled{\ce{^{174}Hf}}{0.364 & 3.273 & {3.268} & 6.695 & {6.688} & 11.114 & {11.097} & 16.398 & {16.337}}
\IsoRowFilled{\ce{^{176}Hf}}{0.392 & 3.288 & {3.284} & 6.765 & {6.755} & 11.306 & {11.293} & 16.785 & {16.765}}
\IsoRowFilled{\ce{^{178}Hf}}{0.395 & 3.289 & {3.291} & 6.771 & {6.785} & 11.321 & {11.356} & 16.816 & {16.850}}
\IsoRowFilled{\ce{^{180}Hf}}{0.463 & 3.311 & {3.306} & 6.883 & {6.868} & 11.643 & {11.617} & 17.507 & {17.480}}
\IsoRowFilled{\ce{^{182}Hf}}{0.434 & 3.304 & {3.294} & 6.843 & {6.814} & 11.527 & {11.473} & 17.252 & {17.184}}
\IsoRowFilled{\ce{^{184}Hf}}{0.382 & 3.283 & {3.264} & 6.741 & {6.696} & 11.240 & {11.198} & 16.650 & {---}}

\hline \hline

\IsoRowFilled{\ce{^{170}W}}{0.101 & 2.953 & {2.950} & 5.555 & {5.587} & 8.701 & {8.700} & 12.351 & {12.133}}
\IsoRowFilled{\ce{^{172}W}}{0.157 & 3.033 & {3.061} & 5.782 & {5.906} & 9.114 & {9.308} & 12.981 & {13.127}}
\IsoRowFilled{\ce{^{174}W}}{0.285 & 3.212 & {3.154} & 6.418 & {6.247} & 10.426 & {10.079} & 15.122 & {14.491}}
\IsoRowFilled{\ce{^{176}W}}{0.287 & 3.214 & {3.215} & 6.426 & {6.458} & 10.445 & {10.524} & 15.157 & {15.222}}
\IsoRowFilled{\ce{^{178}W}}{0.321 & 3.245 & {3.236} & 6.561 & {6.555} & 10.769 & {10.779} & 15.739 & {15.726}}
\IsoRowFilled{\ce{^{180}W}}{0.334 & 3.254 & {3.260} & 6.604 & {6.648} & 10.877 & {10.993} & 15.942 & {16.070}}
\IsoRowFilled{\ce{^{182}W}}{0.427 & 3.301 & {3.289} & 6.832 & {6.796} & 11.493 & {11.427} & 17.177 & {17.095}}
\IsoRowFilled{\ce{^{184}W}}{0.395 & 3.289 & {3.273} & 6.771 & {6.728} & 11.321 & {11.260} & 16.817 & {16.734}}
\IsoRowFilled{\ce{^{186}W}}{0.386 & 3.285 & {3.234} & 6.751 & {6.600} & 11.266 & {11.003} & 16.703 & {16.329}}
\IsoRowFilled{\ce{^{188}W}}{0.266 & 3.191 & {3.071} & 6.333 & {6.085} & 10.233 & {9.957} & 14.789 & {---}}
\hline \hline

\IsoRowFilled{\ce{^{176}Os}}{0.098 & 2.950 & 2.927 & 5.546 & 5.495 & 8.686 & 8.569 & 12.330 & 12.095}
\IsoRowFilled{\ce{^{178}Os}}{0.144 & 3.013 & 3.016 & 5.723 & 5.761 & 9.004 & 9.031 & 12.812 & 12.721}
\IsoRowFilled{\ce{^{180}Os}}{0.203 & 3.106 & 3.093 & 6.016 & 6.018 & 9.565 & 9.514 & 13.688 & 13.375}
\IsoRowFilled{\ce{^{182}Os}}{0.238 & 3.156 & 3.154 & 6.197 & 6.258 & 9.937 & 10.073 & 14.291 & 14.276}
\IsoRowFilled{\ce{^{184}Os}}{0.322 & 3.246 & 3.203 & 6.566 & 6.465 & 10.781 & 10.646 & 15.763 & 15.621}
\IsoRowFilled{\ce{^{186}Os}}{0.295 & 3.222 & 3.166 & 6.460 & 6.335 & 10.524 & 10.361 & 15.296 & 15.082}
\IsoRowFilled{\ce{^{188}Os}}{0.239 & 3.157 & 3.082 & 6.200 & 6.065 & 9.944 & 9.753 & 14.302 & 13.999}
\IsoRowFilled{\ce{^{190}Os}}{0.130 & 2.992 & 2.934 & 5.662 & 5.624 & 8.893 & 8.924 & 12.642 & 12.624}

\hline \hline
\end{longtable}

\clearpage

\scriptsize
\setlength{\LTpre}{0pt}
\setlength{\LTpost}{0pt}

\begin{longtable}{|c|c|c|c|c|c|c|c|c|c|}
\caption{Reduced transition-probability ratios for even--even nuclei in the ground-state band. CBS results are compared with experimental values.}
\\\hline

\label{tab:be2_ratios}
\multirow{3}{*}{Isotope}
& \multirow{3}{*}{$r_\beta$}
& \multicolumn{8}{c|}{Transition ratios} \\ \cline{3-10}
&
& \multicolumn{2}{c|}{$\dfrac{B(E2;4_1^+\!\to2_1^+)}{B(E2;2_1^+\!\to0_1^+)}$}
& \multicolumn{2}{c|}{$\dfrac{B(E2;6_1^+\!\to4_1^+)}{B(E2;2_1^+\!\to0_1^+)}$}
& \multicolumn{2}{c|}{$\dfrac{B(E2;8_1^+\!\to6_1^+)}{B(E2;2_1^+\!\to0_1^+)}$}
& \multicolumn{2}{c|}{$\dfrac{B(E2;10_1^+\!\to8_1^+)}{B(E2;2_1^+\!\to0_1^+)}$}
\\ \cline{3-10}
&
& CBS & Exp
& CBS & Exp
& CBS & Exp
& CBS & Exp \\
\hline
\endfirsthead

\multicolumn{10}{c}{\tablename\ \thetable\ -- continued}\\
\hline
\multirow{3}{*}{Isotope}
& \multirow{3}{*}{$r_\beta$}
& \multicolumn{8}{c|}{Transition ratios} \\ \cline{3-10}
&
& \multicolumn{2}{c|}{$\dfrac{B(E2;4_1^+\!\to2_1^+)}{B(E2;2_1^+\!\to0_1^+)}$}
& \multicolumn{2}{c|}{$\dfrac{B(E2;6_1^+\!\to4_1^+)}{B(E2;2_1^+\!\to0_1^+)}$}
& \multicolumn{2}{c|}{$\dfrac{B(E2;8_1^+\!\to6_1^+)}{B(E2;2_1^+\!\to0_1^+)}$}
& \multicolumn{2}{c|}{$\dfrac{B(E2;10_1^+\!\to8_1^+)}{B(E2;2_1^+\!\to0_1^+)}$}
\\ \cline{3-10}
&
& CBS & Exp
& CBS & Exp
& CBS & Exp
& CBS & Exp \\
\hline
\endhead

\hline
\multicolumn{10}{r}{continued on next page}\\
\endfoot

\hline
\endlastfoot


\IsoRowFilled{\isotope[122]{Ce}}{0.292 & {1.484} & --- & {1.734} & --- & {1.940} & --- & {2.121} & --- }  
\IsoRowFilled{\isotope[124]{Ce}}{0.280 & {1.488} & --- & {1.747} & --- & {1.960} & --- & {2.145} & --- } 
\IsoRowFilled{\isotope[126]{Ce} \cite{IIMURA20221}}{0.210 & {1.524} & 1.341(264) & {1.837} & 0.580(240) & {2.093} & 0.652(119) & {2.310} & ---} 
\IsoRowFilled{\isotope[128]{Ce} \cite{ELEKES2015191}}{0.138 & {1.565} & 1.623(401) & {1.925} & 1.712(487) & {2.209} &  2.252(435)& {2.434} & 2.342(597)} 
\IsoRowFilled{\isotope[150]{Ce} \cite{Basu}}{0.272 & {1.492} & 0.769(390) & {1.756} & {---} & {1.974} & {---} & {2.164} & {---}}
\IsoRowFilled{\isotope[152]{Ce}}{0.364 & {1.459} & {---} & {1.665} & {---} & {1.822} & {---} & {1.962} & {---}}
\IsoRowFilled{\isotope[154]{Ce}}{0.410 & 1.449 &  & 1.635 & {---} & 1.767 & {---} & 1.883 & {---}}
\hline \hline
\IsoRowFilled{\isotope[128]{Nd}}{0.273 & {1.492} & --- & {1.755} & --- & {1.973} & --- & {2.162} &  --- }
\IsoRowFilled{\isotope[130]{Nd}}{0.188 & {1.537} & --- & {1.865} & --- & {2.132} & --- & {2.348} & --- }
\IsoRowFilled{\isotope[150]{Nd} \cite{Basu}}{0.098 & {1.584} & 1.558(43) & {1.959} & 1.776(90) & {2.248} & 1.862(204) & {2.478} & 1.733(105)}
\IsoRowFilled{\ce{^{152}Nd \cite{martin_2013}}}{0.382 & 1.455 & 1.306(99) & 1.652 & $1.260(^{+0.304}_{-0.215})$ & 1.798 & {---} & 1.928 & {---}}
\IsoRowFilled{\ce{^{154}Nd}}{0.424 & 1.446 & {---} & 1.628 & {---} & 1.753 & {---} & 1.862 & {---}}
\IsoRowFilled{\ce{^{156}Nd}}{0.454 & 1.442 & {---} & 1.614 & {---} & 1.727 & {---} & 1.821 & {---}}
\IsoRowFilled{\ce{^{158}Nd}}{0.437 & 1.444 & {---} & 1.621 & {---} & 1.741 & {---} & 1.843 & --- }
\hline \hline
\IsoRowFilled{\isotope[132]{Sm}}{0.283 & {1.487} & --- & {1.743} & --- & {1.954} & --- & {2.139} & --- }
\IsoRowFilled{\ce{^{152}Sm \cite{martin_2013}}}{0.199 & 1.531 &1.445(22) & 1.851 & 1.655(33) & 2.114 & 2.021(35) & 2.328 & $2.166(^{+0.243}_{-0.181})$}
\IsoRowFilled{\ce{^{154}Sm \cite{NICA20252}}}{0.363 & 1.460 & 1.392(42) & 1.666 & 1.642(53) & 1.824 & 1.813(96) & 1.965 & 1.784(96)}
\IsoRowFilled{\ce{^{156}Sm}}{0.443 & 1.443 & {---} & 1.618 & {---} & 1.736 & {---} & 1.835 & {---}}
\IsoRowFilled{\ce{^{158}Sm}}{0.460 & 1.441 & {---} & 1.612 & {---} & 1.723 & {---} & 1.814 & {---}}
\IsoRowFilled{\ce{^{160}Sm}}{0.457 & 1.441 & {---} & 1.613 & {---} & 1.725 & {---} & 1.818 & {---}}
\IsoRowFilled{\ce{^{162}Sm}}{0.448 & 1.443& {---} & 1.616 & {---} & 1.732 & {---} & 1.829 & {---}}
\IsoRowFilled{\ce{^{164}Sm}}{0.394 & 1.452 & {---} & 1.644 & {---} & 1.785 & {---} & 1.908 & {---}}

\hline \hline
\IsoRowFilled{\ce{^{154}Gd \cite{NICA20252}}}{0.203 & 1.529 & 1.542(31) & 1.846 & 1.704(53) & 2.107 & 1.970(78) & 2.320 & $2.236(^{+0.317}_{-0.260})$}
\IsoRowFilled{\ce{^{156}Gd \cite{REICH20122537}}}{0.339 & 1.467 & 1.397(31) & 1.686 & 1.561(50) & 1.859 & 1.693(94) & 2.013 & 1.661(79)}
\IsoRowFilled{\ce{^{158}Gd \cite{NICA20171}}}{0.415 & 1.448 & 1.465(42) & 1.632 & {---} & 1.762 & 1.667(157) & 1.875 & 1.717(157)}
\IsoRowFilled{\ce{^{160}Gd}}{0.441 & 1.444 & {---} & 1.619 & {---} & 1.737 & {---} & 1.837 & {---}}
\IsoRowFilled{\ce{^{162}Gd}}{0.440 & 1.444 & {---} & 1.620 & {---} & 1.738 & {---} & 1.839 & {---}}
\IsoRowFilled{\ce{^{164}Gd}}{0.476 & 1.439 & {---} & 1.606 & {---} & 1.712 & {---} & 1.797 & {---}}
\IsoRowFilled{\ce{^{166}Gd}}{0.455 & 1.442 & {---} & 1.614 & {---} & 1.726 & {---} & 1.820 & {---}}
\hline \hline

\IsoRowFilled{\ce{^{156}Dy \cite{REICH20122537}}}{0.139 & 1.564 & 1.632(24) & 1.923 & 1.760(89) & 2.206 & 1.873(57) & 2.432 & 2.067(201)}
\IsoRowFilled{\ce{^{158}Dy \cite{NICA20171}}}{0.306 & 1.478 & 1.430(86) & 1.718 & 1.828(219) & 1.914 & 1.828(378) & 2.087 & 1.720(271)}
\IsoRowFilled{\ce{^{160}Dy \cite{Nica:2021cny}}}{0.368 & 1.458 & 1.455(60) & 1.662 & $1.216(^{+0.073}_{-0.063})$ & 1.816 & $1.675(^{+0.145}_{-0.129})$ & 1.955 & 1.680(80)}
\IsoRowFilled{\ce{^{162}Dy \cite{Nica:2024ttv}}}{0.416 & 1.448 & 1.430(59) & 1.632 & 1.495(82) & 1.761 & 1.712(88) & 1.873 & $1.732(^{+0.122}_{-0.107})$}
\IsoRowFilled{\ce{^{164}Dy }}{0.430 & 1.445 & {1.351(64)~\cite{PhysRevC.101.024313}} & 1.625 & {1.754(441)~\cite{PhysRevC.101.024313}} & 1.747 & {1.422(67)~\cite{singh_chen_2018}} & 1.853 & {1.697(89)\cite{singh_chen_2018}}}
\IsoRowFilled{\ce{^{166}Dy \cite{PhysRevC.101.024313}}}{0.464 & 1.441 & {---} & 1.610 & {---} & 1.720 & {---} & 1.811 & {---}}
\IsoRowFilled{\ce{^{168}Dy}}{0.461 & 1.441 & {---} & 1.611 & {---} & 1.722 & {---} & 1.813 & {---}}
\IsoRowFilled{\ce{^{170}Dy}}{0.495 & 1.437 & {---} & 1.600 & {---} & 1.701 & {---} & 1.779 & {---}}
\hline \hline
\IsoRowFilled{\ce{^{160}Er \cite{Nica:2021cny}}}{0.233 & 1.512 & 1.426(69) & 1.805 & 1.556(105) & 2.049 & $1.716(^{+0.536}_{-0.360})$ & 2.254 & $1.716(^{+0.536}_{-0.360})$}
\IsoRowFilled{\ce{^{162}Er \cite{Kocheva2026Er}}}{0.323 & 1.472 & $1.351(^{+0.110}_{-0.084})$ & 1.700 & $1.443(^{+0.255}_{-0.191})$& 1.884 & $1.432(^{+0.492}_{-0.293})$ & 2.048 & $0.551(^{+0.119}_{-0.087})$}
\IsoRowFilled{\ce{^{164}Er \cite{singh_chen_2018}}}{0.382 & 1.455 & 1.262(149) & 1.652 & {---} & 1.799 & 1.665(101) & 1.929 & 1.714(97)}
\IsoRowFilled{\ce{^{166}Er \cite{Baglin2008}}}{0.387 & 1.454 & 1.438(61) & 1.649 & 1.705(100) & 1.792 & 1.719(76) & 1.920 & 1.797(89)}
\IsoRowFilled{\ce{^{168}Er \cite{Baglin2010}}}{0.463 & 1.441 & 1.498(51) & 1.611 & 1.991(92) & 1.721 & 1.662(69) & 1.812 & 1.446(67)}
\IsoRowFilled{\ce{^{170}Er \cite{BAGLIN20181}}}{0.467 & 1.440 & {---} & 1.609 & {---} & 1.718 & 1.779(148) & 1.807 & 1.538(110)}
\IsoRowFilled{\ce{^{172}Er}}{0.471 & 1.440 & {---} & 1.608 & {---} & 1.715 & {---} & 1.803 & {---}}
\hline \hline
\IsoRowFilled{\ce{^{162}Yb} \cite{Nica:2024ttv}}{0.036 & 1.598 & 1.559(75) & 1.981 & $1.418(^{+0.102}_{-0.088})$ & 2.274 & $1.856(^{+0.669}_{-0.373})$ & 2.507 & $1.336(^{+0.669}_{-0.372})$}
\IsoRowFilled{\ce{^{164}Yb \cite{singh_chen_2018}}}{0.229 & 1.514 & 1.595(76) & 1.812 & 1.700(83) & 2.058 & 1.970(680) & 2.264 & 1.847(740)}
\IsoRowFilled{\ce{^{166}Yb \cite{Baglin2008}}}{0.323 & 1.472 & 1.424(88) & 1.701 & 1.524(102) & 1.885 & 1.675(227) & 2.049 & 1.623(842)}
\IsoRowFilled{\ce{^{168}Yb \cite{PETKOV2017240}}}{0.354 & 1.462 & 1.478(151) & 1.673 & 1.565(129) & 1.836 & 1.373(94) & 1.982 & 1.225(100)}
\IsoRowFilled{\ce{^{170}Yb}}{0.419 & 1.447 & {1.463(61)~\cite{PhysRevC.95.034316}} & 1.630 & {1.556(328)~\cite{PhysRevC.95.034316}} & 1.758 & {1.800(159)~\cite{BAGLIN20181}} & 1.869 & {1.771(135)~\cite{BAGLIN20181}}}
\IsoRowFilled{\ce{^{172}Yb \cite{Singh1995}}}{0.454 & 1.442 & 1.420(95) & 1.614 & 1.509(142) & 1.727 & 1.887(190) & 1.822 & 1.769(110)}
\IsoRowFilled{\ce{^{174}Yb \cite{Browne1999}}}{0.462 & 1.441 & 1.393(62) & 1.611 & 1.841(255) & 1.721 & 1.930(116) & 1.812 & 1.667(114)}
\IsoRowFilled{\ce{^{176}Yb \cite{basunia_2006}}}{0.453 & 1.442 & 1.475(148) & 1.615 & 1.628(135) & 1.728 & 1.639(280) & 1.823 & 1.749(177)}
\IsoRowFilled{\ce{^{178}Yb}}{0.496 & 1.438 & {---} & 1.600 & {---} & 1.701 & {---} & 1.778 & {---}}
\hline \hline
\IsoRowFilled{\ce{^{166}Hf \cite{Baglin2008}}}{0.136 & 1.566 & 1.578(102) & 1.927 & 1.727(139) & 2.211 & 2.188(263) & 2.437 & ---}
\IsoRowFilled{\ce{^{168}Hf \cite{Baglin2010}}}{0.228 & 1.514 & 1.584(106) & 1.812 & 1.851(144) & 2.059 & 2.273(341) & 2.265 & 2.403(405)}
\IsoRowFilled{\ce{^{170}Hf \cite{BAGLIN20181}}}{0.274 & 1.491 & 1.445(60) & 1.754 & 1.681(85) & 1.971 & 1.890(98) & 2.160 & 2.060(158)}
\IsoRowFilled{\ce{^{172}Hf}}{0.339 & 1.467 & {1.412(103)~\cite{PhysRevC.91.044301}} & 1.686 & $0.912^{+0.418}_{-0.218}$~\cite{PhysRevC.91.044301} & 1.859 & {---} & 2.015 & {---}}
\IsoRowFilled{\ce{^{174}Hf}}{0.364 & 1.459 & {1.454(88)~\cite{PhysRevC.99.024316}} & 1.665 & $1.021(^{+0.356}_{-0.212})$~\cite{PhysRevC.99.024316} & 1.823 & {---} & 1.964 & {---}}
\IsoRowFilled{\ce{^{176}Hf}}{0.392 & 1.452 & {1.381(96)~\cite{PhysRevC.99.024316}} & 1.645 & $1.077^{+0.586}_{-0.282}$~\cite{PhysRevC.91.044301, PhysRevC.99.024316} & 1.786 & {---} & 1.911 & {---}}
\IsoRowFilled{\ce{^{178}Hf}}{0.395 & 1.452 & {$1.816^{+0.327}_{-0.239}$~\cite{PhysRevC.99.024316}} & 1.644 & {1.356(90)~\cite{PhysRevC.99.024316}} & 1.783 & {1.454(45)~\cite{Achterberg2009, PhysRevC.99.024316}} & 1.907 & {1.577(57)~\cite{Achterberg2009, PhysRevC.99.024316}}}
\IsoRowFilled{\ce{^{180}Hf}}{0.463 & 1.441 & {$1.510(^{+0.182}_{-0.150})$\cite{PhysRevC.99.024316}} & 1.611 & {1.452(105)~\cite{PhysRevC.99.024316}} & 1.720 & 1.581(87)~\cite{McCutchan2015, PhysRevC.99.024316} & 1.811 & 1.536(80)~\cite{McCutchan2015, PhysRevC.99.024316}}
\IsoRowFilled{\ce{^{182}Hf}}{0.434 & 1.445 & {---} & 1.622 & {---} & 1.743 & {---} & 1.847 & {---}}
\IsoRowFilled{\ce{^{184}Hf}}{0.382 & 1.455 & {---} & 1.652 & {---} & 1.799 & {---} & 1.929 & {---}}
\hline \hline
\IsoRowFilled{\ce{^{170}W \cite{BAGLIN20181}}}{0.101 & 1.583 & {1.444(149)} & 1.957 & {1.524(119)} & 2.247 & {1.532(405)} & 2.476 & {1.371(324)}}
\IsoRowFilled{\ce{^{172}W \cite{Singh1995}}}{0.157 & 1.555 & {1.433(164)} & 1.904 & {1.520(220)} & 2.182 & {1.696(230)} & 2.405 & {1.579(272)}}
\IsoRowFilled{\ce{^{174}W \cite{Browne1999}}}{0.285 & 1.487 & 1.741(146) & 1.741 & 3.037(626) & 1.951 & 1.778(252) & 2.135 & 1.185(236)}
\IsoRowFilled{\ce{^{176}W \cite{PhysRevC.106.024326}}}{0.287 & 1.486 & {1.518(81)} & 1.739 & {1.800(72)} & 1.948 & {1.253(149)} & 2.130 & {$1.118(^{+0.260}_{-0.178})$}}
\IsoRowFilled{\ce{^{178}W \cite{PhysRevC.106.024326}}}{0.321 & 1.473 & {$1.442(^{+0.089}_{-0.083})$} & 1.703 & {$1.494(^{+0.871}_{-0.417})$} & 1.888 & {---} & 2.053 & {---}}
\IsoRowFilled{\ce{^{180}W \cite{PhysRevC.106.024326}}}{0.334 & 1.468 & 1.447(63) & 1.691 & ${1.411(^{+0.582}_{-0.321})}$ & 1.867 & {---} & 2.025 & {---}}
\IsoRowFilled{\ce{^{182}W \cite{SINGH201521}}}{0.427 & 1.446 & {1.440(76)} & 1.626 & {1.477(163)} & 1.750 & {1.536(134)} & 1.857 & {1.492(141)}}
\IsoRowFilled{\ce{^{184}W \cite{BAGLIN2010275}}}{0.395 & 1.452 & {$1.386(^{+0.046}_{-0.078})$} & 1.644 & {1.511(54)} & 1.783 & 1.544 (47) & 1.907 & $1.578(^{+0.095}_{-0.070})$}
\IsoRowFilled{\ce{^{186}W \cite{BATCHELDER20221}}}{0.386 & 1.454 & $1.281(^{+0.099}_{-0.091})$& 1.649 & $1.610(^{+0.135}_{-0.118})$ & 1.794 & $1.584(^{+0.118}_{-0.109})$ & 1.922 & $1.352(^{+0.161}_{-0.303})$}
\IsoRowFilled{\ce{^{188}W}}{0.266 & 1.495 & {---} & 1.763 & {---} & 1.987 & {---} & 2.178 & {---}}
\hline \hline

\IsoRowFilled{\ce{^{176}Os}}{0.098 & 1.584 & {---} & 1.959 & {---} & 2.249 & {---} & 2.478 & {---}}
\IsoRowFilled{\ce{^{178}Os}}{0.144 & 1.562 & {---} & 1.918 & {---} & 2.200 & {---} & 2.425 & {---}}
\IsoRowFilled{\ce{^{180}Os \cite{McCutchan2015}}}{0.203 & 1.528 & 1.362(229) & 1.846 & 1.135(308) & 2.106 & --- & 2.319 & ---}
\IsoRowFilled{\ce{^{182}Os \cite{PhysRevC.108.024305}}}{0.238 & 1.509 & {---} & 1.799 & {---} & 2.040 & {---} & 2.243 & {---}}
\IsoRowFilled{\ce{^{184}Os \cite{BAGLIN2010275}}}{0.322 & 1.472 & 1.406(402) & 1.701 & {---} & 1.885 & {---} & 2.049 & {---}}
\IsoRowFilled{\ce{^{186}Os \cite{BATCHELDER20221}}}{0.295 & 1.483 & 1.442(81) & 1.730 & 1.976(96)  & 1.933 & 1.410(81) & 2.112 & 2.030(643)}
\IsoRowFilled{\ce{^{188}Os \cite{KONDEV20181}}}{0.239 & 1.509 & 1.716(106) & 1.798 & 1.781(106) & 2.039 & 1.871(119) & 2.241 & $2.387(^{+0.311}_{-0.196})$}
\IsoRowFilled{\ce{^{190}Os \cite{Singh:2020zfc}}}{0.130 & 1.569 & $1.358(^{+0.075}_{-0.051})$ & 1.933 & 1.550(102)& 2.218 & $1.867(^{+0.305}_{-0.237})$ & 2.444 & $ 1.564(^{+0.372}_{-0.304})$}
\hline \hline

\end{longtable}

\clearpage

\scriptsize
\setlength{\LTpre}{0pt}
\setlength{\LTpost}{0pt}

\begin{longtable}{|c|c|c|c|c|c|c|c|}
\caption{Excitation energies (in keV) for even--even nuclei in the \(\beta\)-band (CBS),
including intra-band CBS \(B(E2)\) ratios in the \(\beta\) band.}
\label{tab:beta_band_energies_cbs}\\
\hline
\multirow{2}{*}{Isotope} & \multirow{2}{*}{$r_\beta$}
& \multicolumn{4}{c|}{CBS energies (keV)} & \multicolumn{2}{c|}{CBS \(B(E2)\) ratios} \\
\cline{3-8}
& & $E(0^+_\beta)$ & $E(2^+_\beta)$ & $E(4^+_\beta)$ & $E(6^+_\beta)$
& $\dfrac{B(E2;4^+_\beta\!\to2^+_\beta)}{B(E2;2^+_\beta\!\to0^+_\beta)}$
& $\dfrac{B(E2;6^+_\beta\!\to4^+_\beta)}{B(E2;2^+_\beta\!\to0^+_\beta)}$ \\
\hline
\endfirsthead

\multicolumn{8}{c}{\tablename\ \thetable\ -- continued}\\
\hline
\multirow{2}{*}{Isotope} & \multirow{2}{*}{$r_\beta$}
& \multicolumn{4}{c|}{CBS energies (keV)} & \multicolumn{2}{c|}{CBS \(B(E2)\) ratios} \\
\cline{3-8}
& & $E(0^+_\beta)$ & $E(2^+_\beta)$ & $E(4^+_\beta)$ & $E(6^+_\beta)$
& $\dfrac{B(E2;4^+_\beta\!\to2^+_\beta)}{B(E2;2^+_\beta\!\to0^+_\beta)}$
& $\dfrac{B(E2;6^+_\beta\!\to4^+_\beta)}{B(E2;2^+_\beta\!\to0^+_\beta)}$ \\
\hline
\endhead

\hline
\multicolumn{8}{r}{continued on next page}\\
\endfoot

\hline
\endlastfoot

\ce{^{122}Ce} & 0.292 & 1560.27 & 1724.91 & 2106.97 & 2695.59 & 1.411 & 1.543 \\
\hline
\ce{^{124}Ce} & 0.280 & 1531.83 & 1704.12 & 2103.02 & 2714.19 & 1.410 & 1.544 \\
\hline
\ce{^{126}Ce} & 0.210 & 1390.07 & 1618.60 & 2132.16 & 2876.15 & 1.411 & 1.582 \\
\hline
\ce{^{128}Ce} & 0.138 & 1336.69 & 1652.02 & 2307.36 & 3177.74 & 1.436 & 1.691 \\
\hline
\ce{^{150}Ce} & 0.272 & 1013.52 & 1132.31 & 1406.83 & 1825.70 & 1.410   & 1.545  \\
\hline
\ce{^{152}Ce} & 0.364 & 1292.51 & 1384.15 & 1598.19 & 1934.01 & 1.418 & 1.548 \\
\hline
\ce{^{154}Ce} & 0.410 & 1530.94 & 1615.56 & 1813.34 & 2124.60 & 1.422 & 1.554 \\
\hline \hline

\ce{^{128}Nd} & 0.273 & 1420.43 & 1586.14 & 1969.20 & 2553.97 & 1.410 & 1.545 \\
\hline
\ce{^{130}Nd} & 0.188 & 1212.41 & 1435.49 & 1928.25 & 2624.71 & 1.415 & 1.607 \\
\hline
\ce{^{150}Nd} & 0.098 & 783.20  & 999.81  & 1422.56 & 1963.52 & 1.466 & 1.768 \\
\hline
\ce{^{152}Nd} & 0.382 & 1257.41 & 1338.13 & 1526.75 & 1823.18 & 1.420 & 1.550 \\
\hline
\ce{^{154}Nd} & 0.424 & 1513.46 & 1590.80 & 1771.57 & 2056.18 & 1.422 & 1.556 \\
\hline
\ce{^{156}Nd} & 0.454 & 1684.64 & 1757.06 & 1926.31 & 2192.85 & 1.424 & 1.561 \\
\hline
\ce{^{158}Nd} & 0.437 & 1515.39 & 1587.11 & 1754.73 & 2018.70 & 1.423 & 1.558 \\
\hline \hline
\ce{^{132}Sm} & 0.283 & 1444.96 & 1604.77 & 1975.01 & 2543.16 & 1.410 & 1.543 \\
\hline
\ce{^{152}Sm} & 0.199 & 940.01  & 1103.89 & 1469.12 & 1991.67 & 1.412 & 1.594 \\
\hline
\ce{^{154}Sm} & 0.363 & 1297.57 & 1390.27 & 1606.80 & 1946.46 & 1.418 & 1.547 \\
\hline
\ce{^{156}Sm} & 0.443 & 1789.12 & 1870.99 & 2062.33 & 2363.66 & 1.423 & 1.559 \\
\hline
\ce{^{158}Sm} & 0.460 & 1876.40 & 1954.46 & 2136.87 & 2424.15 & 1.425 & 1.561 \\
\hline
\ce{^{160}Sm} & 0.457 & 1794.46 & 1870.31 & 2047.57 & 2326.73 & 1.424 & 1.561 \\
\hline 
\ce{^{162}Sm} & 0.448 & 1732.68 & 1809.97 & 1990.6 & 2275.06 & 1.424 & 1.560 \\
\hline 
\ce{^{164}Sm} & 0.394 & 1276.74 & 1353.71 & 1533.61 & 1816.53 & 1.421 & 1.552 \\
\hline \hline

\ce{^{154}Gd} & 0.203 & 964.93  & 1129.76 & 1498.22 & 2027.74 & 1.412 & 1.589 \\
\hline
\ce{^{156}Gd} & 0.339 & 1260.60 & 1363.26 & 1602.78 & 1977.12 & 1.416 & 1.544 \\
\hline
\ce{^{158}Gd} & 0.415 & 1625.50 & 1712.72 & 1916.58 & 2237.47 & 1.422 & 1.555 \\
\hline
\ce{^{160}Gd} & 0.441 & 1764.97 & 1846.64 & 2037.51 & 2338.10 & 1.423 & 1.559 \\
\hline
\ce{^{162}Gd} & 0.440 & 1670.68 & 1748.46 & 1930.24 & 2216.51 & 1.423 & 1.558 \\
\hline
\ce{^{164}Gd} & 0.476 & 2054.21 & 2132.19 & 2314.41 & 2601.35 & 1.425 & 1.563 \\
\hline
\ce{^{166}Gd} & 0.455 & 1759.34 & 1834.69 & 2010.78 & 2288.10 & 1.424 & 1.561 \\
\hline \hline

\ce{^{156}Dy} & 0.139 & 893.287 & 1102.88 & 1539.24 & 2119.57 & 1.565 & 1.924 \\
\hline
\ce{^{158}Dy} & 0.306 & 1204.25 & 1321.98 & 1595.76 & 2019.89 & 1.413 & 1.542 \\
\hline
\ce{^{160}Dy} & 0.368 & 1419.58 & 1518.06 & 1748.12 & 2109.21 & 1.418 & 1.548 \\
\hline
\ce{^{162}Dy} & 0.416 & 1663.42 & 1752.12 & 1959.45 & 2285.81 & 1.422 & 1.555 \\
\hline
\ce{^{164}Dy} & 0.430 & 1624.37 & 1704.52 & 1891.86 & 2186.83 & 1.423 & 1.557 \\
\hline
\ce{^{166}Dy} & 0.464 & 2020.17 & 2102.56 & 2295.09 & 2598.30 & 1.424 & 1.562 \\
\hline
\ce{^{168}Dy} & 0.461 & 1954.52 & 2035.28 & 2223.98 & 2521.17 & 1.424 & 1.561 \\
\hline
\ce{^{170}Dy} & 0.495 & 2231.32 & 2307.30 & 2484.82 & 2764.33 & 1.426 & 1.565 \\
\hline \hline

\ce{^{160}Er} & 0.233 & 1122.78 & 1285.58 & 1656.51 & 2206.51 & 1.409 & 1.562 \\
\hline
\ce{^{162}Er} & 0.323 & 1348.14 & 1467.90 & 1746.97 & 2181.58 & 1.414 & 1.542 \\
\hline
\ce{^{164}Er} & 0.382 & 1592.79 & 1695.34 & 1934.95 & 2311.51 & 1.420 & 1.550 \\
\hline
\ce{^{166}Er} & 0.387 & 1447.93 & 1538.43 & 1749.92 & 2082.40 & 1.420 & 1.551 \\
\hline
\ce{^{168}Er} & 0.463 & 2094.64 & 2180.52 & 2381.21 & 2697.28 & 1.425 & 1.562 \\
\hline
\ce{^{170}Er} & 0.467 & 2107.92 & 2192.24 & 2389.27 & 2699.56 & 1.425 & 1.562 \\
\hline
\ce{^{172}Er} & 0.471 & 2106.03 & 2188.57 & 2381.43 & 2685.15 & 1.425 & 1.562 \\
\hline \hline

\ce{^{162}Yb} & 0.036 & 958.60  & 1261.11 & 1809.19 & 2496.81 & 1.505 & 1.840 \\
\hline
\ce{^{164}Yb} & 0.229 & 1097.52 & 1260.63 & 1631.39 & 2178.84 & 1.409 & 1.565 \\
\hline
\ce{^{166}Yb} & 0.323 & 1348.60 & 1468.65 & 1748.37 & 2183.96 & 1.414 & 1.543 \\
\hline
\ce{^{168}Yb} & 0.354 & 1344.84 & 1445.54 & 1680.67 & 2049.11 & 1.418 & 1.546 \\
\hline
\ce{^{170}Yb} & 0.419 & 1761.96 & 1854.45 & 2070.63 & 2410.95 & 1.422 & 1.556 \\
\hline
\ce{^{172}Yb} & 0.454 & 1968.90 & 2053.82 & 2252.28 & 2564.82 & 1.424 & 1.560 \\
\hline
\ce{^{174}Yb} & 0.462 & 2001.93 & 2084.23 & 2276.53 & 2579.39 & 1.425 & 1.561 \\
\hline
\ce{^{176}Yb} & 0.453 & 2046.07 & 2134.91 & 2342.54 & 2669.51 & 1.424 & 1.560 \\
\hline
\ce{^{178}Yb} & 0.509 & 2985.53 & 3072.70 & 3276.28 & 3596.73 & 1.426 & 1.567 \\
\hline \hline

\ce{^{166}Hf} & 0.136 & 1030.19 & 1275.49 & 1783.74 & 2457.16 & 1.438 & 1.695 \\
\hline
\ce{^{168}Hf} & 0.228 & 1092.53 & 1255.43 & 1625.63 & 2171.92 & 1.409 & 1.566 \\
\hline
\ce{^{170}Hf} & 0.274 & 1077.72 & 1202.93 & 1492.40 & 1934.50 & 1.410 & 1.545 \\
\hline
\ce{^{172}Hf} & 0.339 & 1351.58 & 1462.00 & 1719.58 & 2122.14 & 1.416 & 1.544 \\
\hline
\ce{^{174}Hf} & 0.364 & 1454.21 & 1557.74 & 1799.55 & 2178.89 & 1.418 & 1.547 \\
\hline
\ce{^{176}Hf} & 0.392 & 1622.23 & 1720.80 & 1951.16 & 2313.43 & 1.420 & 1.552 \\
\hline
\ce{^{178}Hf} & 0.395 & 1738.57 & 1842.76 & 2086.26 & 2469.24 & 1.421 & 1.552 \\
\hline
\ce{^{180}Hf} & 0.463 & 2450.97 & 2551.22 & 2785.47 & 3154.38 & 1.425 & 1.561 \\
\hline
\ce{^{182}Hf} & 0.434 & 2207.44 & 2313.66 & 2561.94 & 2952.89 & 1.423 & 1.558 \\
\hline
\ce{^{184}Hf} & 0.382 & 1863.02 & 1982.95 & 2263.19 & 2703.60 & 1.420 & 1.550 \\
\hline \hline

\ce{^{170}W}  & 0.101 & 951.74  & 1212.73 & 1724.12 & 2379.65 & 1.464 & 1.763 \\
\hline
\ce{^{172}W}  & 0.157 & 867.83  & 1054.95 & 1454.25 & 1996.93 & 1.426 & 1.656 \\
\hline
\ce{^{174}W}  & 0.285 & 1221.44 & 1355.23 & 1665.28 & 2141.48 & 1.411 & 1.543 \\
\hline
\ce{^{176}W}  & 0.287 & 1229.41 & 1362.63 & 1671.48 & 2146.28 & 1.411 & 1.543 \\
\hline
\ce{^{178}W}  & 0.321 & 1390.55 & 1515.58 & 1806.87 & 2260.25 & 1.414 & 1.542 \\
\hline
\ce{^{180}W}  & 0.334 & 1452.86 & 1574.86 & 1859.40 & 2303.61 & 1.415 & 1.544 \\
\hline
\ce{^{182}W}  & 0.427 & 2174.47 & 2283.55 & 2538.50 & 2939.92 & 1.423 & 1.557 \\
\hline
\ce{^{184}W}  & 0.395 & 2059.69 & 2183.07 & 2471.43 & 2924.98 & 1.420 & 1.552 \\
\hline
\ce{^{186}W}  & 0.386 & 2134.76 & 2269.18 & 2583.29 & 3077.06 & 1.420 & 1.550 \\
\hline
\ce{^{188}W}  & 0.266 & 1432.05 & 1605.82 & 2006.73 & 2616.13 & 1.409 & 1.547 \\
\hline \hline

\ce{^{176}Os} & 0.098 & 805.80  & 1028.78 & 1463.86 & 2020.52 & 1.466 & 1.768 \\
\hline
\ce{^{178}Os} & 0.144 & 887.63  & 1090.98 & 1517.52 & 2088.27 & 1.433 & 1.679 \\
\hline
\ce{^{180}Os} & 0.203 & 1064.15 & 1245.49 & 1651.02 & 2234.10 & 1.411 & 1.589 \\
\hline
\ce{^{182}Os} & 0.238 & 1177.19 & 1343.54 & 1723.42 & 2289.04 & 1.409 & 1.559 \\
\hline
\ce{^{184}Os} & 0.322 & 1561.94 & 1701.30 & 2026.00 & 2531.57 & 1.414 & 1.543 \\
\hline
\ce{^{186}Os} & 0.295 & 1576.73 & 1740.27 & 2119.98 & 2705.77 & 1.412 & 1.542 \\
\hline
\ce{^{188}Os} & 0.239 & 1408.18 & 1606.45 & 2059.36 & 2734.11 & 1.409 & 1.559 \\
\hline
\ce{^{190}Os} & 0.130 & 1199.58 & 1492.31 & 2093.73 & 2885.73 & 1.441 & 1.706 \\
\hline \hline

\end{longtable}

\end{document}